\documentclass[11pt, a4paper]{article}
\usepackage{jcappub}

\usepackage{scalerel}
\usepackage{stackengine,wasysym}

\newcommand\reallywidetilde[1]{\ThisStyle{%
  \setbox0=\hbox{$\SavedStyle#1$}%
  \stackengine{-.1\LMpt}{$\SavedStyle#1$}{%
    \stretchto{\scaleto{\SavedStyle\mkern.2mu\AC}{.5150\wd0}}{.6\ht0}%
  }{O}{c}{F}{T}{S}%
}}

\usepackage{subeqnarray}
\usepackage{lipsum}
\usepackage{accents}
\usepackage{array}
\usepackage[normalem]{ulem}

\usepackage{graphicx}
\usepackage{bm}

\usepackage{pgfplots}
\usepackage{wasysym}
\usepackage{graphicx}

\usepackage{tensor}

\newcommand{\be}{\begin{equation}}
\newcommand{\ee}{\end{equation}}
\newcommand{\bea}{\begin{eqnarray}}
\newcommand{\eea}{\end{eqnarray}}
\newcommand{\beas}{\begin{subeqnarray}}
\newcommand{\eeas}{\end{subeqnarray}}

\newcommand{\vrho}{\varrho}
\newcommand{\bvrho}{\bar{\varrho}}
\newcommand{\dd}{{\rm d}}

\newcommand{\com}[1]{}

\newcommand{\YP}{Y_{\rm P}}

\newcommand{\TNuc}{T_{\rm Nuc}}
\newcommand{\Tcm}{T_{\rm cm}}
\newcommand{\He}[1]{{}^#1{\rm He}}
\newcommand{\Li}{{}^7{\rm Li}}
\newcommand{\Be}{{}^7{\rm Be}}
\newcommand{\had}{\hat{a}^\dagger}
\newcommand{\ha}{\hat{a}}
\newcommand{\hbd}{\hat{b}^\dagger}
\newcommand{\hb}{\hat{b}}
\newcommand{\hcd}{\hat{c}^\dagger}
\newcommand{\hc}{\hat{c}}
\newcommand{\tr}{\mathrm{tr}}
\newcommand{\Tr}{\mathrm{Tr}}
\newcommand{\ddp}[1]{[\dd^3 \vec{p}_{#1}]}
\newcommand{\bnu}{\bar{\nu}}

\newcommand{\Neff}{N_{\mathrm{eff}}}
\newcommand{\vp}{\vec{p}}
\newcommand{\vpp}{{\vec{p} \,}'}
\newcommand{\bi}{{\bar{\imath}}}
\newcommand{\bj}{{\bar{\jmath}}}

\newcommand{\abs}[1]{\left\lvert#1\right\rvert}
\newcommand{\norm}[1]{\left\lVert#1\right\rVert}
\newcommand{\ket}[1]{\lvert #1 \rangle}
\newcommand{\bra}[1]{\langle #1 \rvert}

\definecolor{brickred}{rgb}{0.8, 0.25, 0.33}

\title{Neutrino decoupling including flavour oscillations and primordial nucleosynthesis}

\author[a]{Julien Froustey,}
\author[a]{Cyril Pitrou,}
\author[a,b]{and Maria Cristina Volpe}

\affiliation[a]{Institut d'Astrophysique de Paris, CNRS UMR 7095, Sorbonne Universit{\'e}, 98 bis Bd Arago, 75014 Paris, France}
\affiliation[b]{Astroparticule et Cosmologie, CNRS UMR 7164, Universit{\'e} de Paris, 10 rue Alice Domon et L{\'e}onie Duquet, 75013 Paris, France}

\emailAdd{froustey@iap.fr}
\emailAdd{pitrou@iap.fr}
\emailAdd{volpe@apc.univ-paris7.fr}

\abstract{We revisit the decoupling of neutrinos in the early universe with flavour oscillations. We rederive the quantum kinetic equations which determine the neutrino evolution based on a BBGKY-like hierarchy, and include for the first time the full collision term, with both on- and off-diagonal terms for all relevant reactions.  We focus on the case of zero chemical potential and solve these equations numerically. We also develop an approximate scheme based on the adiabatic evolution in the matter basis. In fact, the large difference between the oscillations and cosmological time scales allows to consider averaged flavour oscillations which can speed up the numerical integration by two orders of magnitude, when combined with a direct computation of the differential system Jacobian. The approximate numerical scheme is also useful to gain more insight into the physics of neutrino decoupling. Including the most recent results on plasma thermodynamics QED corrections, we update the effective number of neutrinos to $N_{\rm eff} = 3.0440$.  Finally we study the impact of flavour oscillations during neutrino decoupling on the subsequent primordial nucleosynthesis.}

\begin{document}
\maketitle
\flushbottom


\section{Introduction}

The hot Big Bang model predicts that several physical phenomena take place when the Universe temperature reaches the MeV scale. Long before the temperature reaches this threshold, the Universe consists in a plasma of coupled photons, electrons, positrons, neutrinos and antineutrinos at equilibrium. But when the temperature drops below $\sim 2 \, \mathrm{MeV}$, weak interactions become too weak to keep (anti)neutrinos in thermal contact with the electromagnetic plasma: neutrinos \emph{decouple} and form the cosmic neutrino background, a key prediction of the standard cosmological model. Soon after, the temperature decreases below the electron mass and $e^\pm$ pairs annihilate into photons, reheating the electromagnetic plasma compared to the bath of neutrinos. If one considers those two events to be well-separated in time, entropy conservation leads to the standard ratio for the temperatures of neutrino and photon backgrounds, $T_\gamma / T_\nu = (11/4)^{1/3}$.

However, the overlap between neutrino decoupling and $e^\pm$ annihilations, known as \emph{incomplete neutrino decoupling}, leads to slightly non thermal neutrino spectra, and to an increased neutrino energy density (both typically of order 1\%), which is usually described by an effective number of thermalised neutrinos $\Neff$ departing from $3$ \cite{Dodelson_Turner_PhRvD1992,Dolgov1992,Dolgov_NuPhB1997,Esposito_NuPhB2000,Mangano2002,Grohs2015,Froustey2019}. An accurate prediction of the neutrino spectra requires to take into account multiple physical effects, including QED radiative corrections to the plasma equation of state \cite{Heckler_PhRvD1994,Mangano2002,Bennett2020}. Furthermore, the small but non-vanishing masses of neutrinos and their mixings are the cause of the famous neutrino oscillations, given that mass eigenstates differ from flavour eigenstates~\cite{GiuntiKim}. Neutrino mixings can thus influence the process of neutrino decoupling, in particular the flavour dependence of spectral distortions. Flavour oscillations have already been included in calculations of neutrino decoupling \cite{Mangano2005,Relic2016_revisited,Gariazzo_2019,Akita2020}, yet approximating some collision terms for computational purposes, either neglecting off-diagonal components or replacing them by damping approximations. An alternative using effective equilibrium spectra for all the species involved \cite{Escudero_2018,Escudero_2020} allowed to obtain accurate results while reducing drastically the computation time. Although useful, this method cannot fully capture the effect induced by flavour oscillations on neutrino spectra.

A robust and precise prediction of the consequences of incomplete neutrino decoupling is crucial since neutrinos impact many cosmological stages:
\begin{enumerate}
	\item During Big-Bang Nucleosynthesis (BBN), neutrinos control neutron/proton conversions as they participate to weak interactions, and the frozen neutron abundance subsequently affects nuclear reactions and light element relics \cite{Pitrou_2018PhysRept,Froustey2019}.
	\item During the Cosmic Microwave Background (CMB) formation, the free streaming of neutrinos is crucial to predict the CMB angular spectrum. Also, the value of $\Neff$ affects the cosmological expansion, and thus also the radiative transfer of CMB. From these effects, CMB alone can be used to place constraints on $\Neff$ ($N_{\rm eff} = 2.99 \pm 0.17$ at 68\% confidence~\cite{Planck18}) or in combination with BBN constraints on primordial light elements~\cite{Pitrou_2018PhysRept}.
	\item In the late universe, neutrino free streaming also affects structure formation, via its effect on the growth of perturbations. This is used to place the constraint $\sum_\nu m_\nu < 0.12 \,{\rm eV}$ (see e.g.~\cite{Planck18}) on the sum of neutrino masses.
\end{enumerate}

It is striking that neutrino masses play a key role in both the earliest stage 1 and the latest stage 3 for very different reasons.
In stage 1, neutrino oscillations, which are due to small neutrino mass-squared differences and mixing angles, affect the non-thermal part of the spectra, as they lead to less distortion in electron-type neutrinos and more distortion in other types than if there were no oscillations at all. Also oscillations lead to a mild modification of $N_{\rm eff}$. In stage 3, and due to cosmological redshifting, all neutrinos undergo at some point a transition from being very relativistic (they behave gravitationally like decoupled photons) to being non-relativistic (they then behave like cold dark matter). This transition depends only on neutrino masses and not on mixing angles, since frozen neutrino spectra inherited from stage 1 are generated incoherently in the mass basis.  
Finally, stage 2 would also be affected beyond the standard cosmological model, if we were to consider exotic physics with increased neutrino self-interactions, so that they would still behave effectively as a perfect fluid around CMB formation~\cite{Kreisch:2019yzn,Grohs:2020xxd}.

This interplay between the various cosmological eras implies that it is crucial to understand neutrino decoupling as precisely as possible, in order to use these predictions as initial conditions for the subsequent eras. For instance, current constraints from CMB on cosmological parameters~\cite{Planck18} were placed using $N_{\rm eff}=3.046$  when solving numerically for the linear evolution of cosmological perturbations.

For stage 1, the inclusion of neutrino masses and mixings requires the numerical solution of the full neutrino quantum kinetic equations (QKEs). Various approaches were used to derive them, e.g., a perturbative expansion of the density matrix \cite{SiglRaffelt}, or the Closed-Time-Path (CTP) formalism for the two-point function \cite{Vlasenko:2013fja,BlaschkeCirigliano}. A hierarchy can be built for the neutrino density matrix, corresponding to a relativistic generalization of the Bogoliubov-Born-Green-Kirkwood-Yvon (BBGKY) equations. This formalism has been applied to derive the most general mean-field equations for astrophysical neutrinos \cite{Volpe_2013, Volpe_2015}, introducing notably neutrino-antineutrino pairing correlations and wrong helicity contributions due to the neutrino mass.

The main goal of this work is to reevaluate the standard value of $N_{\rm eff}$ and the distorted neutrino spectra, including all relevant effects to reach a $10^{-4}$ precision, also including the effect of neutrino masses. To this aim we first derive the neutrino QKEs, extending the work of \cite{Volpe_2013} for astrophysical environments, and implement two-body collisions in an isotropic and homogeneous environment, including neutrino self-interactions with their full matrix structure.\footnote{Note that helicity, or spin coherence \cite{Vlasenko:2013fja,SerreauVolpe}, that requires anisotropy, is not considered in the present work.} Then we numerically solve these QKEs, but also present an approximate solution where an adiabatic evolution is considered, exploiting the different timescales of collisions, mean-field and mixing terms, nearby neutrino decoupling. This procedure allows to maintain the required precision while decreasing substantially the computation time, gaining some physical insight on the role of flavour oscillations in neutrino decoupling. The numerical results we present correspond to the case of zero chemical potential. Finally we investigate the impact of neutrino masses and mixings on BBN predictions, implementing the contribution of the numerical solution of the full QKEs, going beyond works available in the literature~\cite{Mangano2005,Gava:2010kz,Gava_corr}.

The manuscript is structured as follows. The formalism used to determine the neutrino evolution in the early universe is described in section~\ref{sec:BBGKY}, several technical details being gathered in appendices. In section~\ref{sec:ATAO}, the approximate scheme used in computations is presented; whereas results for the key observables (neutrino spectra and $N_{\rm eff}$) are given in section~\ref{sec:Results}. Finally section~\ref{sec:BBN} is devoted to the effect that incomplete neutrino decoupling has on the nucleosynthesis, and to the comparison with previous results \cite{Froustey2019} obtained without taking into account neutrino masses and mixings.  Natural units ($\hbar=c=k_B=1$) are used throughout the manuscript.

\section{Derivation of quantum kinetic equations}
\label{sec:BBGKY}

In this section, we present a derivation from first principles of the neutrino quantum kinetic equations, which generalize the Boltzmann kinetic equation for distribution functions to account for neutrino masses and mixings. We present the BBGKY hierarchy that was historically derived for a non-relativistic $N-$body system and heavily used in nuclear physics \cite{Cassing1990,Reinhard1994,Lac04,Simenel,Lac14}, but that can also be applied to a relativistic system such as neutrinos and antineutrinos in the early universe. We extend the work done in \cite{Volpe_2013}, where the BBGKY formalism was applied to derive extended mean-field equations for astrophysical applications, and include the collision term. Neutrino QKEs were previously derived using different approaches (see e.g.~\cite{SiglRaffelt,BlaschkeCirigliano}).

\subsection{BBGKY formalism}

The exact evolution of a $N-$body system under the Hamiltonian $\hat{H}$ is given by the Liouville-von Neumann equation for the many-body density matrix
\begin{equation}
\label{eq:vonneumann}
i \frac{\dd \hat{D}}{\dd t} = [\hat{H}, \hat{D}] \, ,
\end{equation}
where $\hat{D} = \ket{\Psi}\bra{\Psi}$, with $\ket{\Psi}$ the quantum state, from which we define the $s$-body reduced density matrices, 
\begin{equation}
\label{eq:defrhos}
\hat{\varrho}^{(1 \cdots s)} \equiv \frac{N !}{(N-s)!} \mathrm{Tr}_{s+1\dots N} \hat{D} \, ,
\end{equation}
with components (we drop the superscript $^{(1\cdots s)}$, redundant with the number of indices):
\begin{equation}
\label{eq:defrhos2}
\varrho^{i_1 \cdots i_s}_{j_1 \cdots j_s} \equiv \langle \hat{a}_{j_s}^\dagger \cdots \hat{a}_{j_1}^\dagger \hat{a}_{i_1} \cdots \hat{a}_{i_s} \rangle \, ,
\end{equation}
where the indices $i,j$ label a set of quantum numbers (species $\phi_i$, momentum $\vp_i$, helicity $h_i$) which describe a one-particle quantum state. For instance,
\begin{equation}
\sum_i{\had_i}= \sum_{\phi_i} \sum_{h_i} \int{\ddp{i} \, \had_{\phi_i}(\vp_i,h_i)} \qquad \text{with} \qquad \ddp{i} \equiv \frac{\dd^3 \vp_i}{(2 \pi)^3 2 E_i}   \, .
\end{equation}
The central object is the one-body reduced density matrix \cite{SiglRaffelt},
\begin{equation}
\label{eq:defrho}
\vrho^i_j \equiv \langle \ha_j^\dagger \ha_i \rangle \, ,
\end{equation}
whose diagonal entries correspond to the standard occupation numbers. 

The Hamiltonian for this system is given by the sum of the kinetic and the two-body interaction terms,
\begin{equation}
\hat{H} = \hat{H}_0 + \hat{H}_{\rm int} = \sum_{i,j}{t^{i}_{j} \, \had_i \ha_j} + \frac14 \sum_{i,j,k,l}{\tilde{v}^{ik}_{jl} \, \had_i \had_k \ha_l \ha_j}  \label{eq:defHint} \, .
\end{equation} 
The interaction matrix elements are fully anti-symmetrized by construction:
\begin{equation}
\label{eq:defvint}
\bra{ik} \hat{H}_{\rm int} \ket{jl} \equiv \tilde{v}^{ik}_{jl}= - \tilde{v}^{ki}_{jl} = \tilde{v}^{ki}_{lj} \, .
\end{equation}
This set of definitions ensures proper transformation laws under a unitary transformation $\psi^i = \mathcal{U}^{i}_{a} \psi^a$: all lower indices are covariant while upper indices are contravariant, namely,
\begin{equation}
\label{eq:transfo}
\vrho^a_b = {\mathcal{U}^\dagger}^a_i \, \vrho^i_j \, \mathcal{U}^j_b \quad ; \quad t^a_b = {\mathcal{U}^\dagger}^a_i \, t^i_j \, \mathcal{U}^j_b \quad ; \quad \tilde{v}^{ac}_{bd} = {\mathcal{U}^\dagger}^a_i {\mathcal{U}^\dagger}^c_k \, \tilde{v}^{ik}_{jl} \, \mathcal{U}^j_b \mathcal{U}^l_d \, .
\end{equation}

The evolution equation for $\vrho$ can be obtained directly via the Ehrenfest theorem. One can also apply partial traces to \eqref{eq:vonneumann}, which leads to the well-known \emph{BBGKY hierarchy} \cite{Volpe_2013,Bogoliubov,BornGreen,Kirkwood,Yvon}, whose first two equations read explicitly\footnote{We made explicit the components of the tensors compared to the expressions found in \cite{Volpe_2013} or \cite{Lac04,Simenel}.} (Einstein summation convention implied):
\begin{equation}
\label{eq:hierarchy}
\left \{ \begin{aligned}
i  \frac{\dd \vrho^{i}_{j}}{\dd t} &= \left( t^{i}_{k} \vrho^{k}_{j} - \vrho^{i}_{k} t^{k}_{j} \right) + \frac12 \left(\tilde{v}^{ik}_{ml} \vrho^{ml}_{jk} - \vrho^{ik}_{ml} \tilde{v}^{ml}_{jk} \right)\,,\\
i  \frac{\dd \vrho^{ik}_{jl}}{\dd t} &= \left(t^{i}_{r} \vrho^{rk}_{jl} + t^{k}_{p} \vrho^{ip}_{jl} + \frac12 \tilde{v}^{ik}_{rp} \vrho^{rp}_{jl} - \vrho^{ik}_{rl} t^{r}_{j} - \vrho^{ik}_{jp} t^{p}_{l} - \frac12 \vrho^{ik}_{rp} \tilde{v}^{rp}_{jl} \right)  \\
&\qquad \qquad + \frac12 \left(\tilde{v}^{im}_{rn} \vrho^{rkn}_{jlm} + \tilde{v}^{km}_{pn} \vrho^{ipn}_{jlm} - \vrho^{ikm}_{rln} \tilde{v}^{rn}_{jm} - \vrho^{ikm}_{jpn} \tilde{v}^{pn}_{lm} \right)  \,.\end{aligned} \right.
\end{equation}

More than simply recasting in a less compact form the very complicated problem \eqref{eq:vonneumann}, this hierarchy furnishes a set of evolution equations which depend on higher-order reduced density matrices, and lead to natural truncation schemes for practical applications. The simplest non-trivial closure is the so-called Hartree-Fock or \emph{mean-field} approximation, which corresponds physically to the propagation of particles in a potential due to the interactions with the particles of the background. Mathematically, it consists in neglecting the correlated parts in the two-body density matrix and the higher order density matrices. Separating the uncorrelated and the correlated contributions, the two-body density matrix reads \cite{Lac04,Lac14}
\begin{equation}
\label{eq:splitrho}
\vrho^{ik}_{jl} \equiv 2 \vrho^{i}_{[j} \vrho^{k}_{l]} + C^{ik}_{jl} \equiv \vrho^{i}_{j} \vrho^{k}_{l} - \vrho^{i}_{l} \vrho^{k}_{j} + C^{ik}_{jl} \, .
\end{equation}
Inserting this decomposition into \eqref{eq:hierarchy}, we get:
\begin{equation}
\label{eq:eqvrho}
i  \frac{\dd \vrho^{i}_{j}}{\dd t} = \left( \left[t^{i}_{k} + \Gamma^{i}_{k}\right] \vrho^{k}_{j} - \vrho^{i}_{k} \left[t^{k}_{j} + \Gamma^{k}_{j}\right] \right) + \frac12 \left(\tilde{v}^{ik}_{ml} C^{ml}_{jk} - C^{ik}_{ml} \tilde{v}^{ml}_{jk} \right) 
=  \left [ \hat{t} + \hat{\Gamma} , \hat{\vrho} \right]^{i}_{j} + i \,  \hat{\mathcal{C}}^{i}_{j} \, ,
\end{equation}
where the mean-field potential $\hat{\Gamma}$ is defined as (for once, we make explicit the summation)
\begin{equation}
\label{eq:Gamma}
\Gamma^{i}_{j} = \sum_{k,l}{\tilde{v}^{ik}_{jl} \vrho^{l}_{k}} \, .
\end{equation}
The mean-field approximation then consists in neglecting $C^{ik}_{jl} \simeq 0$ and keeping only the commutator part in \eqref{eq:eqvrho}. However, in the context of neutrino decoupling in the early universe, one seeks a generalization of the Boltzmann equation for neutrino distribution functions \cite{Dolgov_NuPhB1997,Esposito_NuPhB2000,Mangano2002,Grohs2015,Froustey2019}, which describes the evolution of densities under two-body collisions. In other words, we need to truncate the hierarchy \eqref{eq:hierarchy} assuming the \emph{molecular chaos} ansatz: correlations between the one-body density matrices arise from two-body interactions between uncorrelated matrices. This prescribes the form of $C^{ik}_{jl}(t)$, leading to the following formal expression for the collision term (see appendix~\ref{app:derivation_collision} for details):
\begin{multline}
\label{eq:C11}
\mathcal{C}^{i_1}_{i_1'} = \frac14 \left(\tilde{v}^{i_1 i_2}_{i_3 i_4} \vrho^{i_3}_{j_3} \vrho^{i_4}_{j_4} \tilde{v}^{j_3 j_4}_{j_1 j_2} (\hat{1} -\vrho)^{j_1}_{i_1'} (\hat{1} - \vrho)^{j_2}_{i_2} - \tilde{v}^{i_1 i_2}_{i_3 i_4} (\hat{1} -\vrho)^{i_3}_{j_3} (\hat{1} - \vrho)^{i_4}_{j_4} \tilde{v}^{j_3 j_4}_{j_1 j_2} \vrho^{j_1}_{i_1'} \vrho^{j_2}_{i_2} \right. \\ 
\left. + (\hat{1} -\vrho)^{i_1}_{j_1} (\hat{1} - \vrho)^{i_2}_{j_2} \tilde{v}^{j_1 j_2}_{j_3 j_4} \vrho^{j_3}_{i_3} \vrho^{j_4}_{i_4} \tilde{v}^{i_3 i_4}_{i_1' i_2}  - \vrho^{i_1}_{j_1} \vrho^{i_2}_{j_2} \tilde{v}^{j_1 j_2}_{j_3 j_4} (\hat{1} - \vrho)^{j_3}_{i_3} (\hat{1}-\vrho)^{j_4}_{i_4} \tilde{v}^{i_3 i_4}_{i_1' i_2}  \right)\,.
\end{multline}
The collision term has the standard structure ‘‘gain $-$ loss $+$ h.c.'', which will be made more explicit when we give the full expressions for a system of neutrinos and antineutrinos interacting with standard model weak interactions. In \eqref{eq:C11}, the indices $(i_k,j_k)$ will correspond to a definite momentum $\vp_k$.

We will now focus on the case of the early universe and consider three active species of neutrinos in a background of electrons, positrons (and photons). The influence of baryons can be discarded given their negligible density compared to relativistic species (the baryon-to-photon ratio is $\eta \equiv n_b/n_\gamma \simeq 6.1\times 10^{-9}$ from the most recent measurement of the baryon density~\cite{Planck18}).

\subsection{The case of neutrinos in the early universe}

Assuming the universe to be homogeneous and isotropic in the period of interest, the density matrices read,\footnote{The annihilation and creation operators satisfy the equal time anticommutation rules
\[\{\ha_{\nu_\alpha}(\vp, h),\had_{\nu_\beta}(\vpp, h') \} = (2\pi)^3 \, 2E_p \, \delta^{(3)}(\vp - \vpp) \, \delta_{h h'} \, \delta_{\alpha \beta} \ ; \ \{\had_{\nu_\alpha}(\vp, h),\had_{\nu_\beta}(\vpp, h') \} = \{\ha_{\nu_\alpha}(\vp, h),\ha_{\nu_\beta}(\vpp, h') \} = 0 \]
Similar relations hold for the antiparticle operators.}
\begin{align}
\label{eq:homogeneity}
 \langle \had_{\nu_\beta}(\vpp, h') \ha_{\nu_\alpha}(\vp, h)  \rangle &= (2 \pi)^3 \, 2 E_p \, \delta^{(3)}(\vp - \vpp) \delta_{h h'} \, \vrho^\alpha_\beta(p,t) \, \delta_{h-}  \, , \\
  \langle \hbd_{\nu_\alpha}(\vp, h) \hb_{\nu_\beta}(\vpp, h')  \rangle &= (2 \pi)^3 \, 2 E_p \, \delta^{(3)}(\vp - \vpp) \delta_{h h'}  \, \bvrho^\alpha_\beta(p,t) \, \delta_{h+} \, .
\end{align}
The Kronecker delta ensures that only left-handed neutrinos and right-handed antineutrinos are included, whereas wrong helicity contributions can be present in anisotropic environments \cite{Volpe_2015}. The energy function is $E_p = p$ for neutrinos\footnote{We always neglect the small neutrino masses compared to their typical momentum, except for the vacuum term since the diagonal momentum contribution disappears from the evolution equation (section~\ref{subsubsec:vacuum}).} (while it would be $E_p = \sqrt{p^2 + m_e^2}$ for electrons and positrons).  Moreover, in the subspace of charged leptons, the density matrices are diagonal and correspond to the distribution functions $f_e(p,t)$ and $f_{\bar{e}}(p,t)$. 

In the following, we will apply the BBGKY formalism to a system of neutrinos, leaving the inclusion of antineutrinos\footnote{Note that the antineutrino density matrix $\bvrho^i_j \equiv \langle \hb_i^\dagger \hb_j \rangle$ is defined with a transposed convention, compared to the neutrino density matrix, to have similar evolution equations and transformation properties.} to appendix~\ref{app:antiparticles}. Note that, for a relativistic system, the hierarchy is given by an infinite set of equations. The one-body density matrix will be the neutrino one, with the notation $\vrho^\alpha_\beta$ (instead of $\vrho^{\nu_\alpha}_{\nu_\beta}$) for clarity. Furthermore, all quantities being diagonal in momentum space, we only deal with the diagonal values $A(p)$ of operators $A^{\vp}_{\vpp} = A(p) \bm{\delta}_{\vp \vpp}$, where the ‘‘Kronecker symbol'' in momentum space is $\bm{\delta}_{\vp \vpp} = (2 \pi)^3 \, 2 E_p \, \delta^{(3)}(\vp - \vpp)$.

We now calculate the relevant expressions of the vacuum, the mean-field \eqref{eq:Gamma} and collision \eqref{eq:C11} terms for neutrino evolution.

\subsubsection{Vacuum term} 
\label{subsubsec:vacuum}

The neutrino kinetic term is easily calculated in the mass basis, where it is diagonal by definition (the basis elements being the eigenstates of the vacuum Hamiltonian $\hat{H}_0$):
\begin{equation}
\left. t^{a}_{b}(p)\right|_{\text{mass basis}} \simeq p \delta^a_b + \frac{m_a^2}{2p} \delta^a_b  \, .
\end{equation}
Since terms proportional to the identity do not contribute to flavour evolution, the first term will later disappear from the evolution equation. In the flavour basis, the vacuum term is obtained following the transformation laws \eqref{eq:transfo}:
\begin{equation}
t^i_j = p \delta^i_j + \left( U \frac{\mathbb{M}^2}{2p} U^\dagger \right)^i_j \, ,
\end{equation}
with $\mathbb{M}^2$ the matrix of mass-squared differences and $U$ the Pontercorvo-Maki-Nakagawa-Sakata (PMNS) mixing matrix \cite{GiuntiKim}.

\subsubsection{Weak interactions} 

Neutrinos and antineutrinos in the early universe interact with each others and with the electrons and positrons composing the homogeneous and isotropic plasma. The interaction Hamiltonian is thus given by the charged- and neutral-current terms from the standard model of weak interactions, expanded at low energies compared to the gauge boson masses. The different expressions and subsequent interaction matrix elements \eqref{eq:defvint} are displayed in the appendix~\ref{app:matrixelements}. 

\paragraph{Mean-field potential.} With the set of all relevant $\tilde{v}^{ik}_{jl}$, one can compute the mean-field potential from \eqref{eq:Gamma}. This procedure is outlined in \cite{Volpe_2013}, and we just quote here the result:\footnote{The absence of extra complex conjugation on $\bvrho$ compared to \cite{Volpe_2013} is due to the transposed definition of the antineutrino density matrix.}
\begin{multline}
\label{eq:Gamma_potential}
\Gamma^{\alpha}_{\beta} = \sqrt{2} G_F (n_e - n_{\bar{e}}) \delta^{\alpha}_{e} \delta^{e}_{\beta} + \sqrt{2} G_F \left(n_\nu - n_{\bnu}\right)^\alpha_\beta \\ - \frac{2 \sqrt{2} G_F p}{m_W^2}(\rho_e + P_e + \rho_{\bar{e}} + P_{\bar{e}})\delta^{\alpha}_{e} \delta^{e}_{\beta} - \frac{8 \sqrt{2} G_F p}{m_Z^2}\left(\rho_\nu + \rho_{\bnu} \right)^{\alpha}_{\beta} \, .
\end{multline}
The first two terms are the particle/antiparticle asymmetric mean-field potentials arising from the V$-$A Hamiltonian. Expanding the gauge boson propagators to next-to-leading order leads to the symmetric terms proportional to the neutrino momentum $p$. This expression is derived in the flavour basis in which $\delta^{\alpha}_{e}$ is the Kronecker symbol. However it can be directly read in any basis, through the contravariant (covariant) transformation of upper (lower) indices \eqref{eq:transfo}.

 The various thermodynamic quantities involved are
\begin{equation}
\begin{aligned}
n_e &= 2 \int{\frac{\dd^3 \vp}{(2 \pi)^3} f_e(p)}  \\
\left. n_\nu \right|^\alpha_\beta &= \int{\frac{\dd^3 \vp}{(2 \pi)^3} \vrho^\alpha_\beta(p)}
\end{aligned}  \qquad
\begin{aligned} \rho_e  + P_e &= 2 \int{\frac{\dd^3 \vp}{(2 \pi)^3} \left(E_p + \frac{p^2}{3 E_p}\right) f_e(p)}  \\
\left. \rho_\nu \right|^\alpha_\beta &=  \int{\frac{\dd^3 \vp}{(2 \pi)^3} \, p \, \vrho^\alpha_\beta(p)}
\end{aligned} \, ,
\end{equation}
and the corresponding quantities for antiparticles are obtained by replacing $f_e \to f_{\bar{e}}$ and $\vrho^\alpha_\beta \to \bvrho^\alpha_\beta$.

The mean-field potentials up to first order in $1/m_{W,Z}^2$ do not usually take into account the non-relativistic nature of electrons and positrons \cite{SiglRaffelt,Mangano2005,Relic2016_revisited,Gariazzo_2019,Akita2020}. Instead, our expression involves both the energy density and the pressure of charged leptons, as mentioned for instance in \cite{NotzoldRaffelt_NuPhB1988}. As expected, we recover the more common expression in the ultra-relativistic limit $\rho_e + P_e \to (4/3) \rho_e$.

\paragraph{Collision integral.}

The collision term is derived by inserting all possible matrix elements in the general expression \eqref{eq:C11}.
This leads to collision integrals previously derived in \cite{SiglRaffelt,BlaschkeCirigliano}, and progressively included in numerical computations, except for the self-interactions, whose off-diagonal components were approximated by damping terms or discarded \cite{Mangano2005,Gava:2010kz,Gava_corr,Relic2016_revisited,Gariazzo_2019}. In appendix~\ref{app:collision}, we illustrate how our formalism applies by carrying out an explicit derivation for neutrino-neutrino scattering, displaying the full matrix structure of the statistical factor.

\subsubsection{Quantum kinetic equations}

We present here the QKE for $\vrho(p,t)$, obtained from \eqref{eq:eqvrho} after dividing each term by the momentum-conserving function $\bm{\delta}_{\vp \vpp}$ from \eqref{eq:homogeneity}. Moreover, the time derivative $\dd/\dd t$ becomes $\partial/\partial t - H p \, \partial/\partial p$ to account for the expansion of the universe, $H\equiv \dot a/a$ being the Hubble rate, given by Friedmann's equation $H^2 = (8 \pi \mathcal{G}/3) \rho$. The QKEs read:
\begin{equation}
i \left[ \frac{\partial}{\partial t} - H p \frac{\partial}{\partial p}\right] \vrho = \Big[ U \frac{\mathbb{M}^2}{2p}U^\dagger, \varrho \Big] + \sqrt{2} G_F \Big[\mathbb{N}_e + \mathbb{N}_\nu, \varrho \Big] - 2 \sqrt{2} G_F p \Big[ \frac{\mathbb{E}_e + \mathbb{P}_e}{m_W^2} + \frac43 \frac{\mathbb{E}_{\nu}}{m_Z^2},\varrho \Big ] + i \mathcal{I} \label{eq:QKE_rho}
\end{equation}
with the matrices defined in flavour space $\mathbb{N}_e \equiv \mathrm{diag}(n_e - n_{\bar{e}},0,0)$, $\mathbb{N}_\nu \equiv n_\nu - n_{\bar{\nu}}$, $\mathbb{E}_e \equiv \mathrm{diag}(\rho_e + \rho_{\bar{e}},0,0)$ (likewise for $\mathbb{P}_e$), and $\mathbb{E}_\nu \equiv \rho_\nu + \rho_{\bar{\nu}}$. Similarly, the QKEs for the antineutrino density matrix read (cf.~appendix~\ref{app:antiparticles}):
\begin{equation}
i \left[ \frac{\partial}{\partial t} - H p \frac{\partial}{\partial p}\right] \bvrho = - \Big[ U \frac{\mathbb{M}^2}{2p}U^\dagger, \bvrho \Big] + \sqrt{2} G_F \Big[\mathbb{N}_e + \mathbb{N}_\nu, \bvrho \Big] + 2 \sqrt{2} G_F p \Big[ \frac{\mathbb{E}_e + \mathbb{P}_e}{m_W^2} + \frac43 \frac{\mathbb{E}_{\nu}}{m_Z^2},\bvrho \Big ] + i \bar{\mathcal{I}} \label{eq:QKE_rhobar}
\end{equation}
Note that only eq.~\eqref{eq:QKE_rho} will be solved numerically, since we will be focussing on the case of zero chemical potential for which antineutrinos evolve like neutrinos.

The collision term is the sum of the contributions from different physical processes: scattering with charged leptons ($\nu e^\pm\leftrightarrow \nu e^\pm$), annihilation ($\nu \bnu \leftrightarrow e^+ e^-$) and self-interactions (involving only $\nu$ and $\bnu$). The expressions for the processes involving charged leptons are exactly the same as the ones quoted in \cite{Relic2016_revisited} [eqs.~(2.4)--(2.10)], and we do not report them here for brevity. This reference, however, does not contain the full expressions for neutrino self-interactions, derived for instance in \cite{BlaschkeCirigliano}. Our expression for the self-interactions contribution to the collision integral reads:\footnote{It is equivalent with eq.~(96) of ref.~\cite{BlaschkeCirigliano} (one only needs to swap the variables $\vp_3 \leftrightarrow \vp_4$ in the second and fourth terms of \eqref{eq:F_sc_nn}). Our expression makes more explicit the ‘‘gain $-$ loss $+$ h.c.'' structure of this collision term.}
\begin{equation}
\label{eq:C_nn}
\begin{aligned}
\mathcal{I}^{[\nu  \nu]} = &\frac12 \frac{2^5 G_F^2}{2 p_1} \int{[\dd^3 \vec{p}_2] [\dd^3 \vec{p}_3] [\dd^3 \vec{p}_4] (2 \pi)^4 \delta^{(4)}(p_1 + p_2 - p_3 - p_4)} \\
&\Big[ (p_1 \cdot p_2)(p_3 \cdot p_4) F_\mathrm{sc}(\nu^{(1)},\nu^{(2)},\nu^{(3)},\nu^{(4)})  \\
&+ (p_1 \cdot p_4)(p_2 \cdot p_3) \left( F_\mathrm{sc}(\nu^{(1)},\bar{\nu}^{(2)},\nu^{(3)},\bar{\nu}^{(4)}) + F_\mathrm{ann}(\nu^{(1)},\bar{\nu}^{(2)},\nu^{(3)},\bar{\nu}^{(4)}) \right) \Big] \, ,
\end{aligned} 
\end{equation}
with the statistical factors for scattering and annihilation processes:
\begin{multline}
\label{eq:F_sc_nn}
F_\mathrm{sc}(\nu^{(1)},\nu^{(2)},\nu^{(3)},\nu^{(4)}) =  \left[ \varrho_4 (1- \varrho_2) + \Tr(\cdots) \right] \varrho_3 (1-\varrho_1) + (1- \varrho_1) \varrho_3 \left[ (1- \varrho_2) \varrho_4 + \Tr(\cdots)\right]  \\
- \left[ (1- \varrho_4) \varrho_2  + \Tr(\cdots)\right] (1-\varrho_3)  \varrho_1 - \varrho_1  (1-\varrho_3)  \left[\varrho_2(1-\varrho_4)  + \Tr(\cdots)\right]  \, ,
\end{multline}
\begin{multline}
\label{eq:F_sc_nbn}
F_\mathrm{sc}(\nu^{(1)},\bar{\nu}^{(2)},\nu^{(3)},\bar{\nu}^{(4)}) = \left[ (1- \bar{\varrho}_2) \bar{\varrho}_4 + \Tr(\cdots) \right] \varrho_3 (1-\varrho_1) + (1- \varrho_1) \varrho_3 \left[ \bar{\varrho}_4 (1- \bar{\varrho}_2) + \Tr(\cdots)\right]  \\
- \left[ \bar{\varrho}_2 (1-\bar{\varrho}_4) + \Tr(\cdots) \right] (1-\varrho_3) \varrho_1 - \varrho_1 (1-\varrho_3) \left[ (1-\bar{\varrho}_4) \bar{\varrho}_2 + \Tr(\cdots)\right] \, ,
\end{multline}
\begin{multline}
\label{eq:F_ann_nn}
F_\mathrm{ann}(\nu^{(1)},\bar{\nu}^{(2)},\nu^{(3)},\bar{\nu}^{(4)}) = \left[ \varrho_3 \bar{\varrho}_4 + \Tr(\cdots) \right] (1-\bar{\varrho}_2) (1-\varrho_1) + (1- \varrho_1) (1-\bar{\varrho}_2) \left[ \bar{\varrho}_4 \varrho_3 + \Tr(\cdots)\right]  \\
- \left[ (1-\varrho_3) (1-\bar{\varrho}_4) + \Tr(\cdots) \right] \bar{\varrho}_2 \varrho_1 - \varrho_1 \bar{\varrho}_2 \left[ (1-\bar{\varrho}_4) (1-\varrho_3) + \Tr(\cdots)\right] \, ,
\end{multline}
where we chose the more compact notation $\vrho_k = \vrho(p_k)$, and $\Tr(\cdots)$ means the trace of the term in front of it.

\subsection{Reduced set of equations}

The full QKE \eqref{eq:QKE_rho} can be recast in a form more suitable for a numerical resolution. Though neutrino density matrices will deviate from kinetic and chemical equilibrium, electrons and positrons undergo very efficient electromagnetic interactions with the photon background, ensuring that their distribution function remains a Fermi-Dirac one at the photon temperature $T_\gamma$ \cite{Grohs2019}.  Due to a very low baryon-to-photon ratio $\eta$, the difference between the electron and positron number densities is very small compared to the number density of relativistic species (e.g.~photons or neutrinos). When electrons and positrons are still relativistic, this implies that their chemical potentials can be safely ignored as they are of the same order as $\eta$. When they annihilate at temperatures lower than the electron mass, the number density difference remains constant leading to a complete asymmetry when positrons have disappeared, and thus to a sizeable chemical potential for electrons, see e.g.~figure~30 of ref.~\cite{Pitrou_2018PhysRept}. However the relic number density of electrons is of the order of $\eta$ and their effect on neutrino decoupling can be completely ignored. We will thus neglect the chemical potential of $e^\pm$ since its effect on neutrino decoupling would be of the order of $\eta$. In addition, we also neglect any CP violating phase in the PMNS matrix (its effect can be handled separately, see end of section~\ref{subsec:results}) or CP breaking reactions, implying that the equality $\vrho = \bvrho$ will be ensured at all times \cite{Gava:2010kz,Gava_corr}. Therefore, we will only solve the equation \eqref{eq:QKE_rho}, in which the antisymmetric mean-field $\mathbb{N}_e + \mathbb{N}_\nu$ vanishes. 

Moreover, since deviations from the equilibrium distribution $\vrho \propto \mathbb{I}$ are small (cf. numerical results below), the mean-field term proportional to $\mathbb{E}_\nu$ will be very close to the identity (because we are at zero chemical potential), so it will give a negligible contribution within the commutator. We thus discard this term in the numerical resolution.  

The most time consuming part of the QKE is the computation of the collision term. Thanks to the homogeneity and isotropy of the early Universe, and the particular form of the scattering amplitudes, the nine-dimensional collision integrals can be reduced to two-dimensional ones \cite{Hannestad_PhRvD1995,Semikoz_Tkachev,Dolgov_NuPhB1997,Grohs2015}.  We follow here the reduction method of ref.~\cite{Dolgov_NuPhB1997}.  Finally, we define the comoving temperature $\Tcm \propto a^{-1}$ \cite{Grohs2015}, which corresponds to the physical temperature of all species when they are strongly coupled, i.e. $T_\nu = T_\gamma = \Tcm$ when $\Tcm \gg 1 \, \mathrm{MeV}$. From this proxy for the scale factor, we define the comoving variables \cite{Esposito_NuPhB2000,Mangano2005}
\begin{equation}
x=m_e/\Tcm\, , \qquad y=p/\Tcm\, ,\qquad z=T_\gamma/\Tcm\, ,
\end{equation}
which are respectively the reduced scale factor, the comoving momentum, and the dimensionless photon temperature, such that $\vrho(p,t)$ is now expressed $\vrho(x,y)$. We also introduce the dimensionless thermodynamic quantities $\bar{\rho} \equiv (x/m_e)^4 \rho $ and $\bar{P} \equiv  (x/m_e)^4 P$.

Therefore, the QKE is rewritten:
\begin{equation}
\frac{\partial \vrho(x,y_1)}{\partial x} = - \frac{i}{xH} \frac{x}{m_e} \left[ U \frac{\mathbb{M}^2}{2y_1}U^\dagger, \varrho \right]  +  i \frac{2 \sqrt{2} G_F}{xH} y_1 \left(\frac{m_e}{x}\right)^5 \left[ \frac{\bar{\mathbb{E}}_e + \bar{\mathbb{P}}_e}{m_W^2} ,\varrho \right ] + \frac{1}{xH} \mathcal{I} \, , \label{eq:QKE_final}
\end{equation}
with the two-dimensional collision integral\footnote{We integrated out the energy delta-function via $\int{p_4 \dd p_4 \, \delta(E_1+E_2-E_3-E_4)} = E_1+E_2-E_3$, since $p_4 \dd p_4 = E_4 \dd E_4$. In \eqref{eq:I_full} $E_4$ stands for $E_1 + E_2 - E_3$.}  (recall that we assume $f_e = f_{\bar{e}}$, which regroups some terms):

\begin{equation}
\label{eq:I_full}
\begin{aligned}
\mathcal{I} = &\frac{G_F^2}{2 \pi^3 y_1} \left(\frac{m_e}{x}\right)^5 \int{y_2 \dd y_2 \, y_3 \dd y_3 \, \bar{E}_4 \times \frac12} \\
\times &\Big[ 4 \left[ 2 d_1 + 2 d_3 + d_2(1,2) + d_2(3,4) - d_2(1,4) - d_2(2,3) \right] \\
&\qquad \qquad  \times \left(F_\mathrm{sc}^{LL}(\nu^{(1)},e^{(2)},\nu^{(3)},e^{(4)}) + F_\mathrm{sc}^{RR}(\nu^{(1)},e^{(2)},\nu^{(3)},e^{(4)})\right) \\
&- 4 x^2 \left[d_1 - d_2(1,3) \right]/\bar{E}_2 \bar{E}_4 \times \left(F_\mathrm{sc}^{LR}(\nu^{(1)},e^{(2)},\nu^{(3)},e^{(4)}) + F_\mathrm{sc}^{RL}(\nu^{(1)},e^{(2)},\nu^{(3)},e^{(4)}) \right) \\
+ \, &4 \left[ d_1 + d_3 - d_2(1,4) - d_2(2,3) \right] \times  \left( F_\mathrm{ann}^{LL}(\nu^{(1)},\bar{\nu}^{(2)},e^{(3)},e^{(4)}) + F_\mathrm{ann}^{RR}(\nu^{(1)},\bar{\nu}^{(2)},e^{(3)},e^{(4)}) \right)  \\
&+ 2 x^2\left[d_1 + d_2(1,2) \right]/\bar{E}_3 \bar{E}_4 \times \left(F_\mathrm{ann}^{LR}(\nu^{(1)},\bar{\nu}^{(2)},e^{(3)},e^{(4)}) + F_\mathrm{ann}^{RL}(\nu^{(1)},\bar{\nu}^{(2)},e^{(3)},e^{(4)}) \right) \\
+ \, &  \left[ d_1 + d_3 + d_2(1,2) + d_2(3,4) \right] \times F_\mathrm{sc}(\nu^{(1)},\nu^{(2)},\nu^{(3)},\nu^{(4)})  \\
&+ \left[ d_1 + d_3 - d_2(1,4) - d_2(2,3) \right] \times \left( F_\mathrm{sc}(\nu^{(1)},\bar{\nu}^{(2)},\nu^{(3)},\bar{\nu}^{(4)}) + F_\mathrm{ann}(\nu^{(1)},\bar{\nu}^{(2)},\nu^{(3)},\bar{\nu}^{(4)}) \right) \Big]
\end{aligned}
\end{equation}
The $d-$functions are $d_i = (x/m_e) d_i^{\rm DHS}$, with $d_i^{\rm DHS}$ defined in \cite{Dolgov_NuPhB1997} as functions of the momenta $p$, hence the prefactor $x/m_e$. Note that \cite{Relic2016_revisited} use a different convention (4 times greater $D-$functions and opposite sign for $D_2$).

In addition to the QKEs, the remaining dynamical equation is the energy conservation equation $\dot{\rho} = - 3 H (\rho + P)$, rewritten as an equation on $z(x)$ \cite{Mangano2002,Bennett2020}. See appendix~\ref{app:QED} for the complete expression including QED corrections to the plasma equation of state.

\section{Adiabatic transfer of averaged oscillations}
\label{sec:ATAO}

Solving the full QKE \eqref{eq:QKE_final} is a priori a considerable numerical challenge because of the need to resolve numerically both the effect of the mean-field terms and of computationally expensive collision integrals. However, previous numerical results \cite{Mangano2005,Relic2016_revisited} seem to indicate that the expected oscillations are somehow ``averaged'' while there is a comparatively slow evolution due to collisions. 

\begin{figure}[!ht]
	\centering
	\includegraphics[width=13cm]{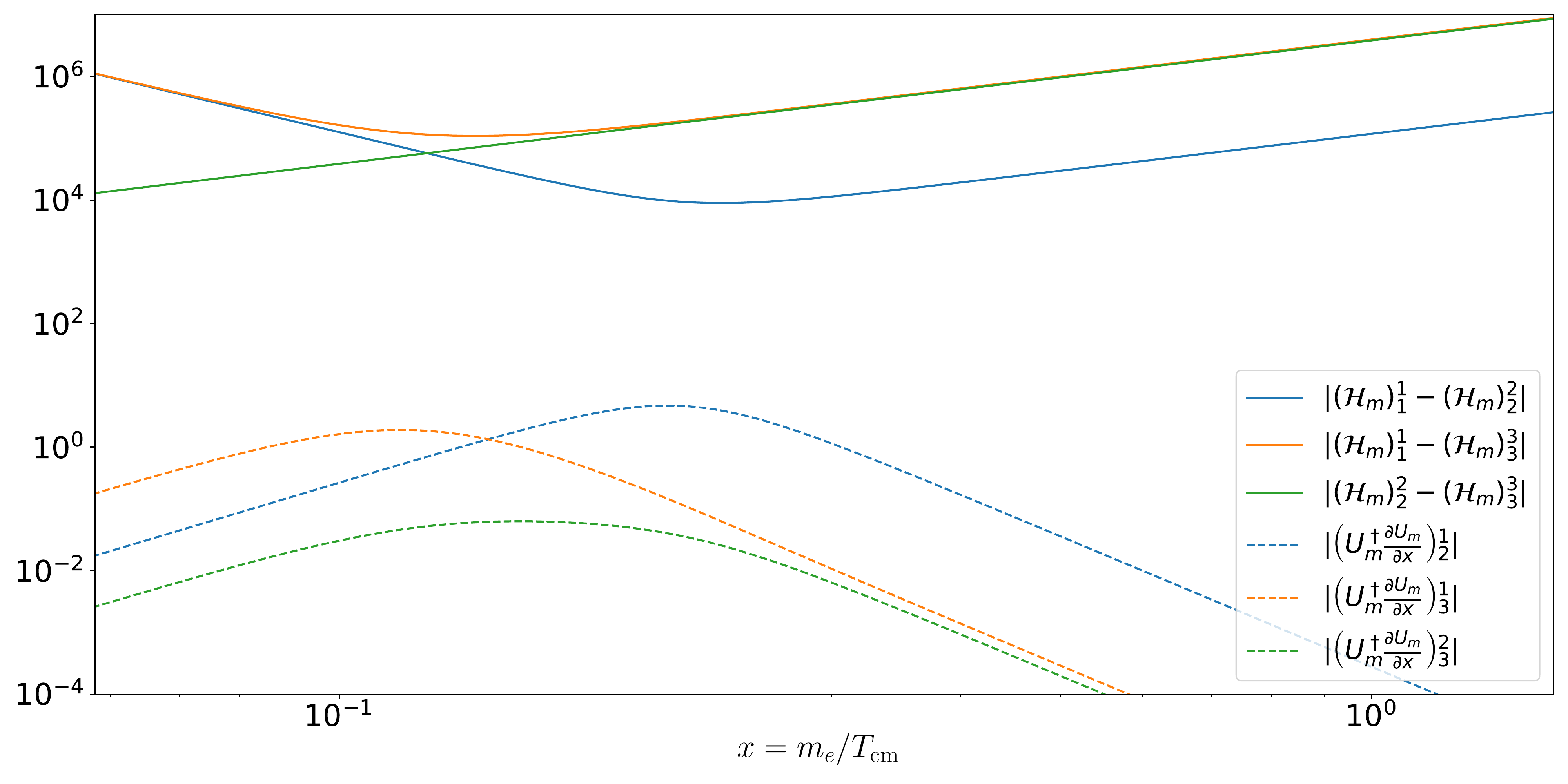}
	\caption{\label{fig:ODG_adiab} Evolution of the different quantities appearing in \eqref{eq:QKEmatt} in the normal hierarchy of masses, for a comoving momentum $y=5$. The condition \eqref{eq:adiab_cond} is satisfied throughout the evolution.}
\end{figure}

\begin{figure}[!ht]
	\centering
	\includegraphics[width=13.5cm]{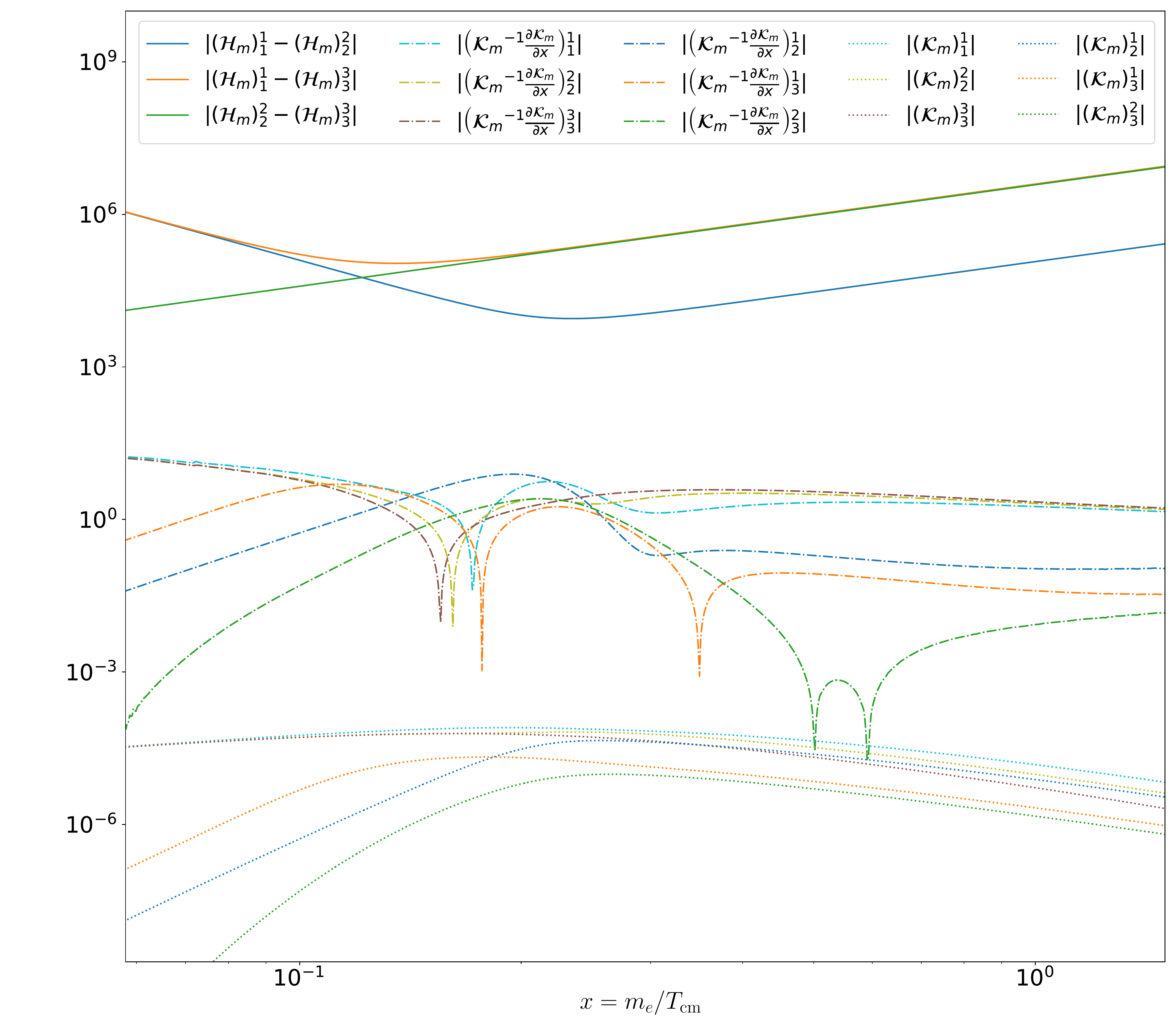}
	\caption{\label{fig:ODG_coll} Comparison of the evolution of the collision term, its relative variation and the effective oscillation frequencies in the normal hierarchy of masses, for a comoving momentum $y=5$. We check that condition \eqref{eq:collis_cond} is satisfied with several orders of magnitude.}
\end{figure}

We thus expect a clear separation of time-scales to hold, allowing for an effective description which correctly captures the salient features of the dynamical evolution. For convenience, let us rewrite the QKE \eqref{eq:QKE_final} in the compact form:
\begin{equation}
\label{eq:QKE_compact}
\frac{\partial \vrho}{\partial x} = - i [\mathcal{H},\vrho] + \mathcal{K} \, ,
\end{equation}
with
 \begin{equation}\label{eq:Hreduced}
 \mathcal{H} \equiv \frac{1}{x H} \left[ \frac{x}{m_e} U \frac{\mathbb{M}^2}{2y}U^\dagger - 2 \sqrt{2} G_F y \left(\frac{m_e}{x}\right)^5 \frac{\bar{\mathbb{E}}_e + \bar{\mathbb{P}}_e}{m_W^2} \right]
\end{equation}
and $\mathcal{K} \equiv \frac{1}{xH} \mathcal{I}$ . 
We treat the $y$ dependence of $\mathcal{H}$ implicitly, as the following procedure must be applied for each $y$. Since the mean-field Hamiltonian $\mathcal{H}$ is Hermitian, it can be diagonalized by the unitary transformation
\begin{equation}
\mathcal{H} = U_m \mathcal{H}_m  U_m^\dagger \qquad \text{with} \qquad (\mathcal{H}_m)^j_k = (\mathcal{H}_m)^j_j \, \delta^{j}_{k} \, .
\end{equation}
The density matrix in the matter basis reads $\vrho_m = U_m^\dagger \, \vrho \, U_m$, and evolves according to
\begin{equation}
\label{eq:QKEmatt}
\frac{\partial \vrho_m}{\partial x} = -i [\mathcal{H}_m,\vrho_m] -  \left[U_m^\dagger \frac{\partial U_m}{\partial x},\vrho_m \right] + U_m^\dagger \mathcal{K} U_m \, .
\end{equation}
The first approximation that we consider is the \emph{adiabatic approximation} \cite{HannestadTamborra,GiuntiKim} which consists in neglecting the time evolution of the matter PMNS matrix compared to the inverse effective oscillation frequency:\footnote{More specifically, we need to check that $\abs{\left(U_m^\dagger \frac{\partial U_m}{\partial x}\right)^j_k} \ll \abs{(\mathcal{H}_m)^j_j - (\mathcal{H}_m)^k_k}$.}
\begin{align}
&\text{\textbf{Adiabatic approximation}}&   \norm{U_m^\dagger \frac{\partial U_m}{\partial x}} &\ll \norm{\mathcal{H}_m} \, . \label{eq:adiab_cond}
\\
  \intertext{This condition means that the effective mixing matrix elements vary very slowly compared to the effective oscillation frequencies, so that the matter basis evolves adiabatically. Such adiabaticity condition is particularly important in presence of Mikheev-Smirnov-Wolfenstein (MSW) resonances \cite{MSW_W,MSW_MS}. Note that the sign of the mean-field contribution to $\mathcal{H}$ \eqref{eq:Hreduced} is opposite to the one encountered due to charged-current neutrino-electron scattering at lowest order, important for astrophysical environments (Sun, supernovae, binary neutron star mergers).
We numerically checked (figure~\ref{fig:ODG_adiab}) that the condition \eqref{eq:adiab_cond} is indeed satisfied throughout the range of temperatures of interest. \endgraf
If we now assume that many oscillations take place before the collision term varies substantially and write the collision term in matter basis $\mathcal{K}_m \equiv U_m^\dagger \mathcal{K} U_m$, its variation frequency $\sim \mathcal{K}_m^{-1} (\partial \mathcal{K}_m/\partial x)$ must be small compared to the effective oscillation frequency $\mathcal{H}_m$. We also assume that the collision rate itself is small compared to the oscillation frequencies, namely} 
&\text{\textbf{Averaged oscillations}}&  \norm{\mathcal{K}_m}, \norm{\mathcal{K}_m^{-1} \frac{\partial \mathcal{K}_m}{\partial x}} &\ll \norm{\mathcal{H}_m} \label{eq:collis_cond}
\, .
\end{align}
If this new separation of time-scales holds (see figure~\ref{fig:ODG_coll}), we can \emph{average} the evolution over many oscillations (the collision term produces at constant rate neutrinos with random initial phases). The non-diagonal parts will then be washed out if the collision rate is not too strong. More precisely, we can write
\begin{equation}
\label{eq:solveapprox}
(\vrho_m)^j_k(x,y) \equiv e^{- i (\mathcal{H}_m)^j_j x} R^{j}_{k}(x,y) e^{i (\mathcal{H}_m)^k_k x} \ \implies \ \frac{\partial R^{j}_{k}}{\partial x} = e^{i (\mathcal{H}_m)^j_j x}(\mathcal{K}_m)^j_k e^{- i (\mathcal{H}_m)^k_k x} \, , 
\end{equation}
where we also assumed a slow variation of $\mathcal{H}_m$, as a consequence of the adiabatic approximation. If \eqref{eq:collis_cond} holds, $\partial R^{j}_{k}/\partial x$ is integrated over many oscillations and the non-diagonal parts vanish.\footnote{As it is sometimes stated, one could phrase it by saying that the off-diagonal terms are washed out by going to a {\it comoving} frame.} This leaves us with the effective equation in matter basis:
\begin{equation}
\label{eq:ATAO}
\text{\textbf{Adiabatic Transfer of Averaged Oscillations}}  \qquad \left\{ \begin{aligned}
 \frac{\partial \tilde{\vrho}_m}{\partial x} &= \reallywidetilde{U_m^\dagger \mathcal{K} U_m} \\
 \vrho_m &= \tilde{\vrho}_m \end{aligned} \right. \, ,
\end{equation}
where the tilde means that we only keep the diagonal terms of $\vrho_m$, then convert it to the flavour basis to compute the collision term $\mathcal{K}$ and only keep the diagonal part of the collision term $U_m^\dagger \mathcal{K} U_m$ when transforming back to the matter basis.

In the flavour basis, the density matrix $\vrho = U_m \tilde{\vrho}_m U_m^\dagger$ has non-diagonal components, while $\tilde{\vrho}_m$ is diagonal. Therefore the collision term destroys the coherence between these components (since it aims at a diagonal $\vrho$ in flavour space, with equilibrium distributions), which modifies the diagonal values of $\vrho_m$ (whose non-diagonal terms average out). 

For clarity, we refer to this approximate numerical scheme to determine the neutrino evolution ``Adiabatic Transfer of Averaged Oscillations'' (ATAO) and  we solve \eqref{eq:ATAO} instead of \eqref{eq:QKE_compact}. 

In the following section, we will numerically solve the QKEs in both the full case and the ATAO approximation and discuss the validity of the approximate numerical solution.

\section{Numerical results}
\label{sec:Results}

\subsection{Numerical implementation}

We integrate numerically the QKE for neutrinos~\eqref{eq:QKE_final}, or \eqref{eq:ATAO} in the ATAO approximation, along with the energy conservation equation~\eqref{eq:zQED}. We use our own code {\tt NEVO} (Neutrino EVOlver) written in Python with the {\tt scipy} and {\tt numpy} libraries.\footnote{Time consuming functions are compiled with the just-in-time compiler {\tt numba}.}

\paragraph{Solver and initial conditions.}

The collision term consists most of the time in nearly compensating gain and loss terms, and for energies larger than $0.1\,{\rm MeV}$, the system is very stiff. Hence, one must rely on an implicit method. We chose the {\tt LSODA} method which consists in a {\tt BDF} method (with adaptative order and adaptative step) when the system is stiff, which switches to an explicit method when not stiff (the {\tt Adams} method). It was first distributed within the {\tt ODEPACK} \cite{ODEPACK} {\tt Fortran} library, but we used the Python wrapper {\tt solve\_ivp} distributed with the Python {\tt scipy} module. We remarked that when setting the absolute and relative error tolerances to $10^{-n}$, the spectra are typically obtained with precision better than $10^{-n+2}$, in agreement with section B.5 of \cite{Gariazzo_2019}. Hence we fixed these error tolerances to $10^{-7}$ so as to obtain results with numerical errors below $10^{-5}$.

The initial common temperature of all species, that is all types of neutrinos and the electromagnetic plasma, is inferred from the conservation of total entropy. Choosing the initial comoving temperature $T_{{\rm cm},{\rm in}}  = 20 \,{\rm MeV}$, the initial common temperature of all species is slightly larger because of early $e^\pm$ annihilations, and given by $T_{\rm in} = z_{\rm in} T_{{\rm cm},{\rm in}}$ with $z_{\rm in}-1 = 7.42\times 10^{-6}$. Had we chosen to start at $T_{{\rm cm},{\rm in}}  = 10 \,{\rm MeV}$, the initial comoving temperature would be $z_{\rm in}-1 = 2.98\times 10^{-5}$, in agreement with Refs.~\cite{Mangano2005,Dolgov_NuPhB1999}. 
As initial condition for the density matrix we take
\begin{equation}
\label{eq:initial_condition}
\vrho(x_{\rm in},y) = \begin{pmatrix}
f_\nu^{({\rm in})}(y) & 0 & 0 \\
0 & f_\nu^{({\rm in})}(y) & 0 \\
0 & 0 & f_\nu^{({\rm in})}(y)
\end{pmatrix} \quad , \quad \text{with} \quad f_\nu^{({\rm in})}(y) \equiv \frac{1}{e^{y/z_{\rm in}} + 1} \, .
\end{equation}

\paragraph{Momentum grid.} 

The neutrino spectra are sampled with $N$ points on a grid in the reduced momentum $y$. When choosing a linear grid, we use the range $0.01 \leq y\leq 16 + [N/20]$, and integrals are evaluated with the Simpson method. However, for functions which decay exponentially for large $y$, it is motivated to use the Gauss-Laguerre quadrature which was already proposed in~\cite{Gariazzo_2019}, and we confirm that this method typically requires half of the grid points to reach the same precision as the one obtained with a linear spacing. In practice, when choosing the nodes and weights of the quadrature, we restrict to $y\leq 20+[N/5]$. When using $N=80$, we have thus restricted nodes to $y \leq 36$, and we used Laguerre polynomials of order $439$ to compute the weights with eq.~(B.14) of \cite{Gariazzo_2019}. Since the tools provided in {\tt numpy} are restricted to much lower polynomial orders, we used {\it Mathematica} to precompute once for all in a few hours the nodes and weights. The results reported in this paper were performed with $N=80$ and the Gauss-Laguerre quadrature, checking that with $N=100$ the differences are smaller than the desired precision.

For each momentum $y_i$ of the grid, and with $N_{\nu}$ flavours, each density matrix has $N_{\nu}^2$ independent degrees of freedom ($N_\nu(N_{\nu} +1)$ real parts and $N_\nu(N_{\nu} -1)$ imaginary parts). In practice we reorganize these independent matrix entries into a vector $A^j(y_i)$ with $j=1,\dots,N_{\nu}^2$ and we concatenate them with the $y_i$ spanning the momentum grid. We thus solve for serialized variables, that is a giant vector of length $N N_\nu^2$. When using the ATAO approximation, one needs only to keep the diagonal part in the matter basis, and the giant vector is of size $N N_\nu$.\footnote{Results are then only converted at the very end in the flavour basis if desired.} Note that we do not store the binned density matrix components $\vrho^\alpha_\beta(y_i)$, which would be sub-optimal. Indeed, if neutrinos decoupled instantaneously, their distribution function would then be
\begin{equation}
\label{eq:fnueq}
f_{\nu}^{\rm (eq)}(x,y) \equiv \frac{1}{e^{y}+1} \, .
\end{equation}
Therefore, we can parametrize the density matrix $\vrho^\alpha_\beta(x,y) = \left[\delta^{\alpha}_{\beta}+a^\alpha_\beta(x,y)\right]\times f_\nu^{\rm (eq)}(x,y)$, and we store the values of $a^\alpha_\beta$, which encapsulate the deviation from instantaneous decoupling.

\paragraph{Numerical optimization via Jacobian computation.} The implicit method requires to solve algebraic equations and thus to obtain the Jacobian of the differential system. For the sake of this discussion, and to alleviate the notation, we ignore the different flavours and consider that we have only one neutrino flavour with spectrum $f(y)$. Noting the grid points $y_i$ and the values of the spectra $f_i = f(y_i)$ on the grid, the differential system is of the type $\partial_x f_i = C_i(x, f_j)$. The implicit method requires the Jacobian $J_{ij} \equiv \partial C_i/\partial f_j$. If no expression is provided, it is evaluated by finite differences in the $\{ f_i\}$ at a given $x$. Since the collision term involves a two-dimensional integral for each point of the grid, its computation on the whole grid is of order ${\cal O}(N^3)$. Hence the computation of the Jacobian with finite differences is of order ${\cal O}(N^4)$. Since algebraic manipulations (mostly the LU decomposition) are at most of order ${\cal O}(N^3)$, reducing the cost of the Jacobian numerical evaluation is crucial to improve the speed of the implicit method. Fortunately, it is possible to compute the Jacobian with an ${\cal O}(N^3)$ complexity. To use a simple example, let us only consider the contribution from the loss part of the neutrino self-interactions, without including Pauli-blocking factors. This component of the collision term, once computed numerically with a quadrature, is of the form
\begin{equation}
\label{ContriC}
C_i(x, f_j) = -\sum_{j,k} w_j w_k g(y_i,y_j,y_k) f_i f_j\,.
\end{equation}
In this expression $\sum_j w_j$ (resp. $\sum_k w_k$) accounts for the integration on $y_2$ (resp. $y_3$) in \eqref{eq:I_full} using a quadrature, and the function $g$ takes into account the specific form of the factor multiplying the statistical function (which is for the contribution considered $f_i f_j$).
Noting then that
\begin{equation}
\partial f_i /\partial f_j = \delta_{ij} \, ,\label{dfdf}
\end{equation}
the Jacobian associated with the contribution \eqref{ContriC} is
\begin{equation}\label{Jimtwocontrib}
J_{i m} = \partial C_i/\partial f_m = -\delta_{im} \sum_{j,k} w_j w_k g(y_i,y_j,y_k) f_j - \sum_{k} w_m w_k g(y_i,y_m,y_k) f_i \, . 
\end{equation}
The complexity of the second sum is of order ${\cal O}(N)$, and since the Jacobian has $N^2$ entries, it leads to a complexity of order ${\cal O}(N^3)$. The first term is not worse even though the double sum is of order ${\cal O}(N^2)$, because it concerns only the diagonal entries of the Jacobian due to the prefactor $\delta_{im}$. More generally for all contributions to the collision term, the complexity when computing the associated Jacobian is always of order ${\cal O}(N^3)$, even when taking into account Pauli-blocking factors which bring terms which are cubic or quartic in the density matrix. For instance, terms similar to \eqref{ContriC}, but with factors $f_i f_j f_k$, are handled with the same method and would lead to three contributions instead of two in \eqref{Jimtwocontrib}. As for terms with factor $f_i f_j f_l$, they would be handled using total energy conservation $y_i + y_j = y_k + y_l$, which allows for instance to replace the variables of summations (e.g. $\sum_{j,k} \to \sum_{j,l}$) when varying with respect to $f_l$. Following these arguments, one notices that the exponent of the complexity for both the collision term and its associated Jacobian is given by the number of independent momenta magnitudes, given that integrations on momenta directions have all been removed with the integration reduction method using the isotropy of momentum distribution. In the case at hand, we have only two-body collisions, for which total energy conservation implies that only three momenta magnitudes are independent, hence the complexity in ${\cal O}(N^3)$. 
When restoring the fact that we do not have a single flavour but density matrices, the discussion is similar when using the serialized variables described above, and again the complexity is of order ${\cal O}(N^3)$. In practice, we found that it takes roughly five times more time to compute a Jacobian than a collision term. Hence, when compared with the finite difference method, providing a numerical method for the Jacobian leads to a factor $N/5$ speed-up. Note that we must also integrate $z$ with eq.~\eqref{eq:zQED} jointly with the density matrices, so that we must pad the Jacobian obtained with the previous description with one extra line and one extra column. Again, the corresponding entries can be deduced using \eqref{dfdf} and their computation is also of order ${\cal O}(N^3)$.
It is worth mentioning that providing a method for the Jacobian is not specific to the ATAO approximation. Indeed, when solving the full QKE one can also compute the Jacobian of the collision term, and one only needs to add the contribution from the vacuum and mean field commutators whose complexity is simply of order ${\cal O}(N^2)$.

When compared with the full QKE method, the ATAO numerical resolution allows to gain at least a factor 5 in time. Hence when using both a method for the Jacobian and the ATAO approximation, we gain typically a factor $N$ and computations that would otherwise last days on CPU clusters, are reduced to just few hours on a single CPU. Moreover, nothing prevents the computation of collision terms and Jacobians to be parallelized on the momentum grid, as we checked on the 4 or 8 CPUs of desktop machines, reducing even further the computation time.

\subsection{Oscillation parameters}
For the numerical calculations, we employ the standard parametrization of the PMNS matrix which reads \cite{Gariazzo_2019,GiuntiKim}
\begin{equation}
\label{eq:PMNS}
U = R_{23} R_{13} R_{12} = \begin{pmatrix} 
c_{12} c_{13} & s_{12} c_{13} & s_{13} \\
- s_{12}c_{23} - c_{12}s_{23}s_{13} & c_{12} c_{23} - s_{12}s_{23}s_{13} & s_{23} c_{13} \\
s_{12}s_{23} - c_{12}c_{23}s_{13} & -c_{12}s_{23} - s_{12}c_{23}s_{13} & c_{23} c_{13}
\end{pmatrix} \, ,
\end{equation}
with $c_{ij} = \cos{\theta_{ij}}$, $s_{ij}=\sin{\theta_{ij}}$ and $\theta_{ij}$ the mixing angles. $R_{ij}$ is the real rotation matrix of angle $\theta_{ij}$ in the $i$-$j$ plane, namely, $(R_{ij})^i_i = (R_{ij})^j_j = c_{ij}$, $(R_{ij})^k_k = 1$ where $k \neq i,j$, $(R_{ij})^i_j = - (R_{ij})^j_i = s_{ij}$ and the other components are zero. Note that we do not introduce yet a CP violating phase, postponing its treatment to appendix~\ref{app:CP}. We use the most recent values from the Particle Data Group \cite{PDG}:
\begin{align}
\left(\frac{\Delta m_{21}^2}{\rm 10^{-5} \, eV^2},\frac{\Delta m_{31}^2}{\rm 10^{-3} \, eV^2},s_{12}^2,s_{23}^2,s_{13}^2\right)_{\rm NH} &= \left(7.53, 2.53, 0.307, 0.545, 0.0218 \right) \, .
\end{align}

For completeness, we also give the most recent values of the physical constants used \cite{PDG}: the Fermi constant $G_F = 1.1663787 \times 10^{-5} \, \mathrm{GeV^{-2}}$ and the gravitational constant $\mathcal{G} = 6.70883 \times 10^{-39} \, \mathrm{GeV^{-2}}$.

\subsection{Neutrino temperature and spectra}
\label{subsec:results}

\begin{figure}[!ht]
	\centering
	\includegraphics[width=12.2cm]{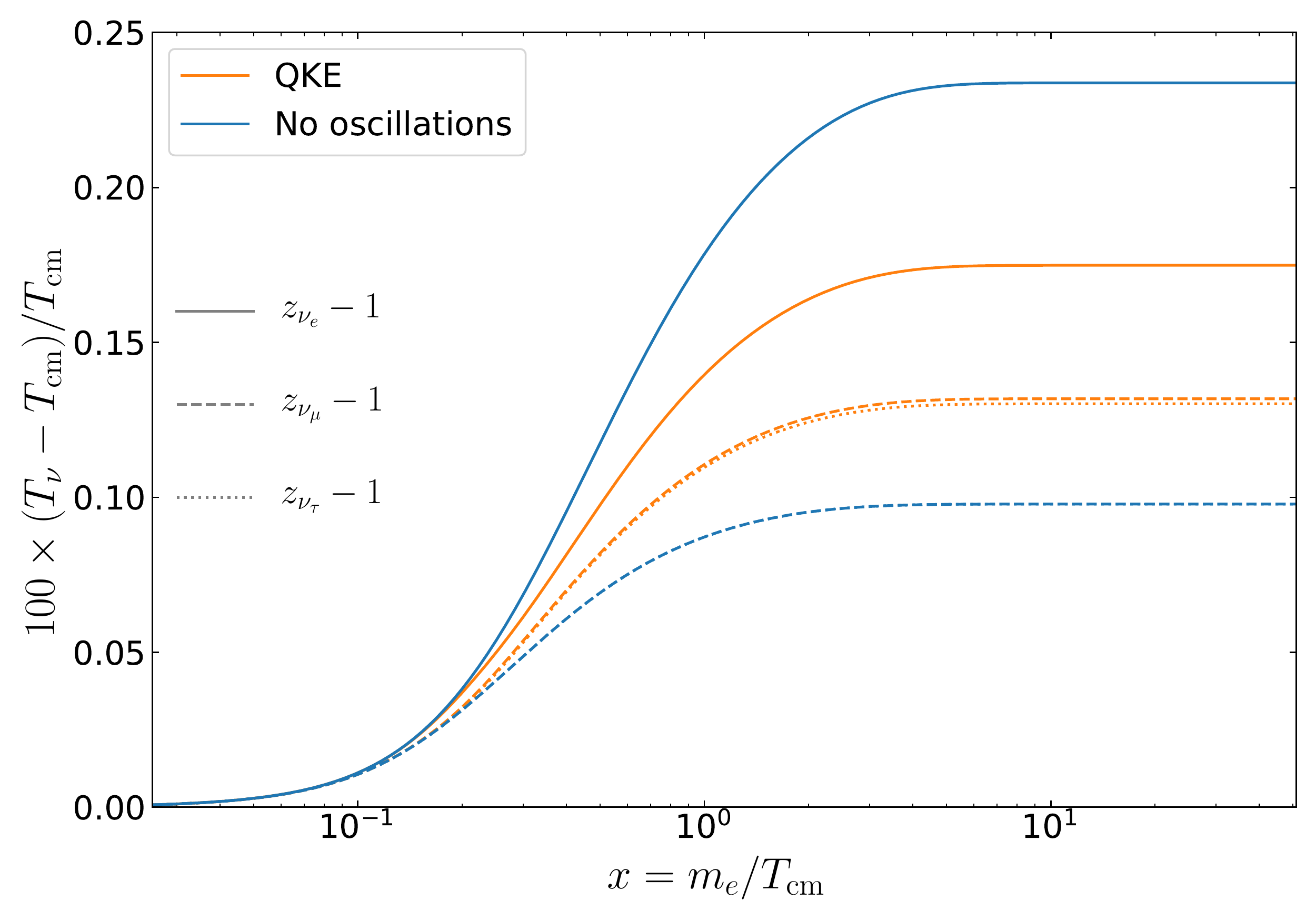}
	\caption{\label{fig:Tnu} Evolution of the effective neutrino temperatures, with and without oscillations. Long before decoupling, they remain equal to the photon temperature $z$, before freezing-out at different values depending on the interaction with the electromagnetic plasma. Without mixing, the distribution function (and thus, the effective temperatures) are identical for $\nu_\mu$ and $\nu_\tau$.}
\end{figure}

\begin{figure}[!ht]
	\centering
	\includegraphics[width=12.2cm]{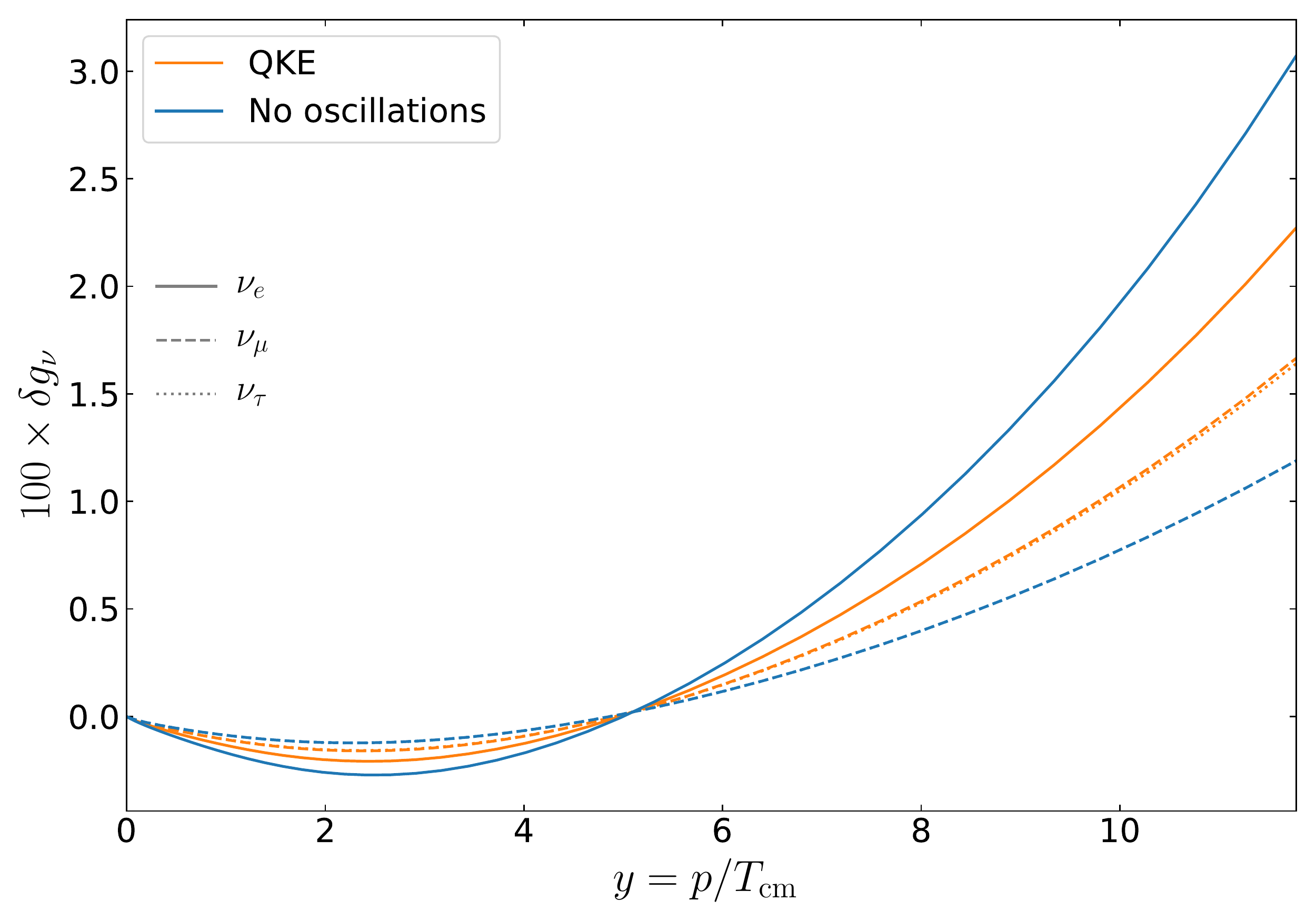}
	\caption{\label{fig:deltagnu} Frozen-out effective spectral distortions, with and without oscillations, for $x_f \simeq 51$ (corresponding to $T_{{\rm cm},f} = 0.01 \, \mathrm{MeV}$). The full QKE results are indistinguishable from the ATAO approximate ones.}
\end{figure}

A convenient parametrization of neutrino spectral distortions consists in separating effective temperatures and residual distortions \cite{Froustey2019}, namely,
\begin{equation}
\label{eq:param_rho}
\vrho^\alpha_\alpha(x,y) \equiv \frac{1}{e^{y/z_{\nu_\alpha}} + 1} \left[1 + \delta g_{\nu_\alpha}(x,y)\right] \, ,
\end{equation}
where the reduced effective temperature $z_{\nu_\alpha} \equiv T_{\nu_\alpha}/\Tcm$ is the reduced temperature of the Fermi-Dirac spectrum with zero chemical potential which has the same energy density as the real distribution:
\begin{equation}
\bar{\rho}_{\nu_\alpha} \equiv \frac78 \frac{\pi^2}{30} z_{\nu_\alpha}^4 \,.
\end{equation}
We plot in figure~\ref{fig:Tnu} the evolution of the neutrino effective temperatures, with and without flavour oscillations. The higher values for the electronic flavour are due to the charged-current processes (that do not exist for muon and tau neutrinos), which increase the transfer of entropy from electrons and positrons. Likewise, the non-thermal residual distortions are more important for $\vrho^e_e$ (see figure~\ref{fig:deltagnu}). This increased energy density of neutrino species has historically been parametrized through the effective number of neutrino species $\Neff$, i.e., the number of instantaneously decoupled neutrino species that would give the same energy density. Long after decoupling, this reads:
\begin{equation}
\label{eq:defNeff}
\rho = \left[1 + \frac78 \left(\frac{4}{11}\right)^{4/3} \Neff \right] \rho_\gamma \ \iff \ \Neff \equiv \left[ \frac{(11/4)^{1/3}}{z} \right]^4 \times \left(z_{\nu_e}^4 + z_{\nu_\mu}^4 + z_{\nu_\tau}^4 \right) \, .
\end{equation}

The final values of the comoving temperatures and $\Neff$ are given in table~\ref{Table:Res_NuDec}. The inclusion of QED corrections to the plasma equation of state up to ${\cal O}(e^3)$ order reduces $\Neff$ by $\sim 0.001$, as predicted in ref.~\cite{Bennett2020}, and already observed in \cite{Akita2020}. Indeed, without these corrections, but keeping the ones at order ${\cal O}(e^2)$, we get $\Neff \simeq 3.0444$ (no oscillations), compared to $\Neff \simeq 3.0434$ with the corrections up to ${\cal O}(e^3)$. 

Flavour oscillations reduce the discrepancy between the different flavours, thus $z_{\nu_e}$ is reduced while $z_{\nu_\mu}$ and $z_{\nu_\tau}$ are increased, with a very slightly higher value for $z_{\nu_\mu}$. This enhanced entropy transfer towards $\nu_\mu$ compared to $\nu_\tau$ is due to the more important $\nu_e-\nu_\mu$ mixing (cf.~figure~\ref{fig:ATAO} and the corresponding discussion).

\renewcommand{\arraystretch}{1.2}

\begin{table}[!htb]
	\centering
	\begin{tabular}{|l|ccccc|}
  	\hline 
  Final values & $z$ & $z_{\nu_e}$  & $z_{\nu_\mu}$ & $z_{\nu_\tau}$ &$\Neff$  \\
  \hline \hline
  Instantaneous decoupling, no QED & $1.40102$ &  $1.00000$ & $1.00000$ &$1.00000$ & $3.00000$ \\ \hline
  No oscillations (NO), QED ${\cal O}(e^3)$ &$1.39800$ & $1.00234$ &  $1.00098$ & $1.00098$ &$3.04340$ \\ \hline
  NO, post-averaging, QED ${\cal O}(e^3)$ & $1.39800$ & $1.00173$ & $1.00130$ & $1.00127$ & $3.04340$   \\
w/o mean-field, QED ${\cal O}(e^3)$ & $1.39796$ & $1.00175$ & $1.00132$ & $1.00131$ & $3.04407$   \\ \hline 
    ATAO, QED ${\cal O}(e^3)$ & $1.39797$ & $1.00175$ & $1.00132$ & $1.00130$ & $3.04397$   \\
    Full QKE, QED ${\cal O}(e^3)$ & $1.39797$ & $1.00175$ & $1.00132$ & $1.00130$ & $3.04397$   \\ \hline
\end{tabular}
	\caption{Frozen-out values of the dimensionless photon and neutrino temperatures, and the effective number of neutrino species. The values without oscillations differ from \cite{Froustey2019} because of the inclusion of QED corrections at order ${\cal O}(e^3)$ in this work (cf.~appendix~\ref{app:QED}). $\Neff$ is different between the ATAO approximation and full QKE calculations at order $10^{-6}$, which we attribute mainly to numerical errors. The implementations in the third and fourth lines are discussed in section~\ref{subsec:transfer}. The post-averaging result corresponds to eq.~\eqref{eq:DefPost}.
	\label{Table:Res_NuDec}}
\end{table}

The deviation of the dimensionless temperatures with respect to $1$ can be expressed as a relative change in the energy density, $\delta \bar{\rho}_\nu = 4 (z_\nu - 1)$. Our values for the increase in the neutrino energy density are $\delta \bar{\rho}_{\nu_e} \simeq  0.70 \, \%$, $\delta \bar{\rho}_{\nu_\mu} \simeq  0.53 \, \%$ and $\delta \bar{\rho}_{\nu_e} \simeq  0.52 \, \%$. This is in agreement with the results of ref.~\cite{Relic2016_revisited} (table 1) or ref.~\cite{Akita2020} (table 2), except for the relative variation of muon and tau flavours: these works obtain a higher reheating of $\nu_\tau$ compared to $\nu_\mu$, while we find the opposite. This is due to a difference in the values of the mixing angles.\footnote{For instance, the older values used in \cite{Mangano2005} lead to higher distortions for $\nu_\mu$ than for $\nu_\tau$.} Nevertheless, if we use the mixing angles from \cite{Relic2016_revisited}, we obtain $\delta \bar{\rho}_{\nu_e} \simeq  0.694 \, \%$, $\delta \bar{\rho}_{\nu_\mu} \simeq  0.525 \, \%$ and $\delta \bar{\rho}_{\nu_\tau} \simeq  0.530 \, \%$. Furthermore, if  ${\cal O}(e^3)$ QED corrections are not included and only the diagonal components of the self-interaction collision term are kept, the spectra reach less flavour equilibration and the results of \cite{Relic2016_revisited} are recovered (at the level of a few $10^{-5}$): $\delta \bar{\rho}_{\nu_e} \simeq  0.706 \, \%$, $\delta \bar{\rho}_{\nu_\mu} \simeq  0.515 \, \%$ and $\delta \bar{\rho}_{\nu_\tau} \simeq  0.522 \, \%$.

Finally, the results in table~\ref{Table:Res_NuDec} show the striking accuracy of the ATAO approximation, as expected since the conditions \eqref{eq:adiab_cond} and \eqref{eq:collis_cond} are satisfied by several orders of magnitude (figures~\ref{fig:ODG_adiab} and \ref{fig:ODG_coll}). The frozen-out values of the comoving temperatures and of $\Neff$ differ by $10^{-6}$, which is beyond our desired accuracy, and beyond the expected effect of neglected contributions.\footnote{Higher order QED corrections to the plasma thermodynamics or subdominant log-dependent contributions are not expected to modify $\Neff$ above order $10^{-5}$ \cite{Bennett2020}.} 

The numerical solution of the QKE shows a larger $\Neff$ value (table~\ref{Table:Res_NuDec}) compared to the no-oscillation case. To understand this slight increase of the total energy density of neutrinos,  one should keep in mind that electron-positron annihilations, which is the dominant process during decoupling, are more efficient in producing electronic type neutrinos (because of the existence of charged-current processes). Now the mixing and mean-field terms tend to depopulate $\nu_e$ and populate the other flavours, which frees some phase space for the reactions which create $\nu_e$, while increasing the effect of Pauli-blocking factors for reactions creating $\nu_{\mu,\tau}$. Since the former are the dominant reactions, the net effect is a larger entropy transfer from $e^\pm$, hence the larger value of $\Neff$. In the next section, we further clarify the effect of mixing and mean-field terms in the light of the ATAO approximation.

To conclude, we find that the value of $\Neff$ predicted by the Standard model of cosmology, including flavour oscillations and QED radiative corrections to the plasma equation of state, is $\Neff=3.0440$ with at least $10^{-4}$ precision. There is one remaining physical ingredient that could modify the value at this order: QED radiative corrections to the collision rates \cite{Passera_QED,Esposito_QED} were estimated to decrease $\Neff$ by $0.001$ \cite{Escudero_2020}. The inclusion and analysis of these corrections are however outside the scope of this paper.

\paragraph{Sensitivity to the parameters of the PMNS matrix.}

The experimental uncertainties on the values of the mixing angles \cite{PDG} lead to small variations of the neutrino distribution functions and $\Neff$. The numerical sensitivity of $\Neff$ to the variation of the mixing angles around their preferred values are:
\begin{equation}
\frac{\partial \Neff}{\partial \theta_{12}} \simeq 1.2 \times 10^{-3} \ \mathrm{rad^{-1}} \quad ; \quad  
\frac{\partial \Neff}{\partial \theta_{13}} \simeq 3.4 \times 10^{-3} \ \mathrm{rad^{-1}} \quad ; \quad
 \abs{\frac{\partial \Neff}{\partial \theta_{23}}} \ll  \abs{\frac{\partial \Neff}{\partial \theta_{12}}},  \abs{\frac{\partial \Neff}{\partial \theta_{13}}} \, .
\end{equation} 
The sensitivity with respect to $\theta_{23}$ is much smaller than for the other mixing angles, and cannot be separated from numerical noise. Given the uncertainties on the mixing angles \cite{PDG}, we estimate the associated variation of $\Neff$ to be $\Delta \Neff \sim 2 \times 10^{-5}$, beyond our accuracy goal. 

Moreover, we neglected up to now a CP violating phase in the PMNS matrix, while some experiments favour a value different from $\delta = 0$ or $\delta = \pi$ \cite{PDG}. Under the assumptions of this work, it can be analytically shown that introducing such a phase would leave $\Neff$ unchanged and only affect the $\nu_\mu$ and $\nu_\tau$ distributions, thus having no effect on BBN (see section~\ref{sec:BBN}). The corresponding results are gathered in appendix~\ref{app:CP}.

\subsection{ATAO transfer functions}
\label{subsec:transfer}

\begin{figure}[ht]
	\centering
     \includegraphics[width=0.98\textwidth]{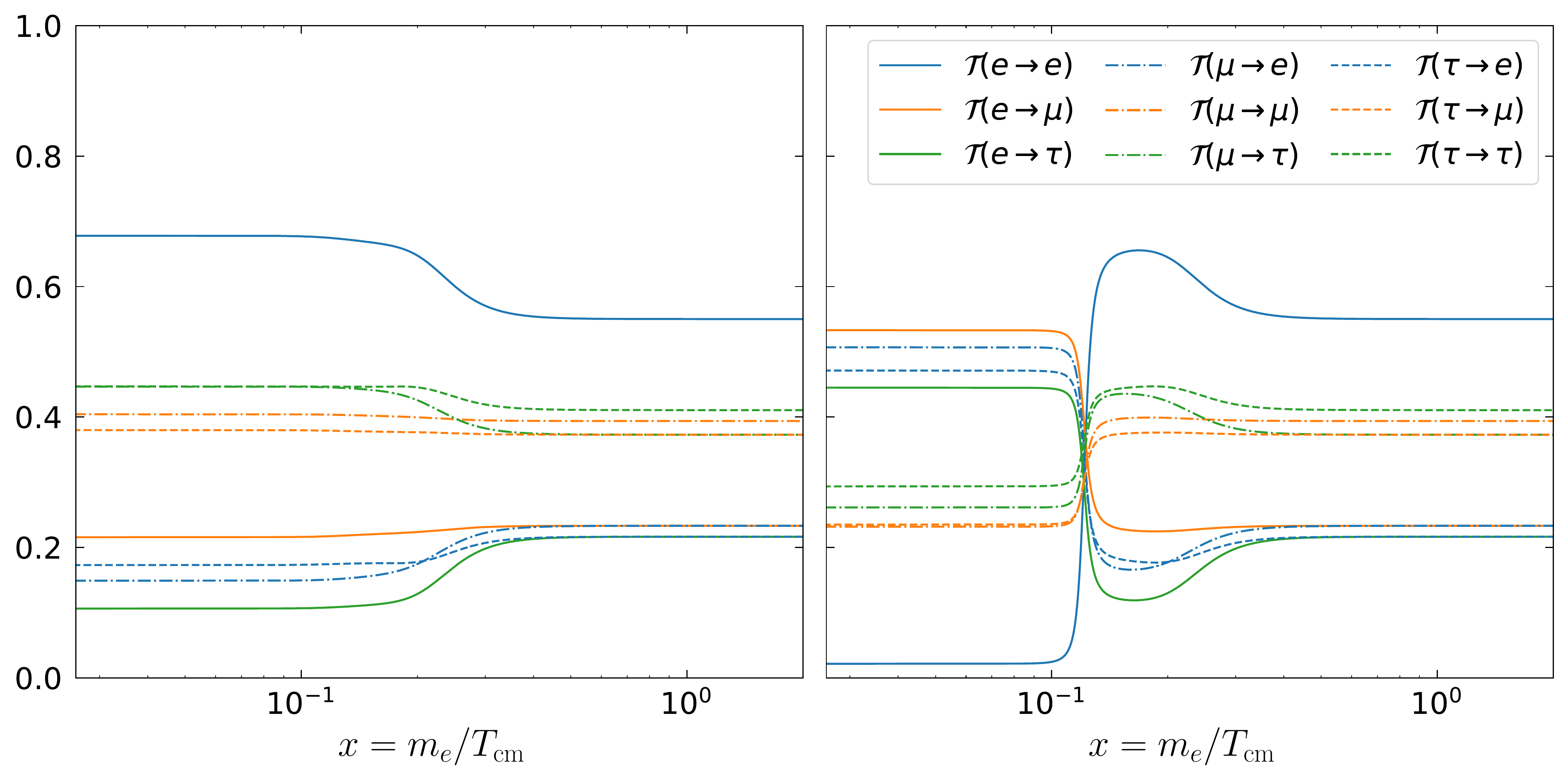}
	\caption{\label{fig:ATAO} Asymptotic ATAO transfer function $\mathcal{T}(\alpha \to \beta, x, y)$ for $y = 5$. {\it Left }: Normal hierarchy. {\it Right }: Inverted hierarchy. The asymptotic values for large $x$ correspond to the vacuum oscillation averages \eqref{VacuumAverage}.}
      \end{figure}

The ATAO approximation allows to get some insight on the impact of the mixings and mean-field terms on the spectral modifications and on $\Neff$. To this purpose, let us define the ATAO transfer function 
\begin{equation}
\label{eq:defATAO}
\mathcal{T}(\alpha \to \beta, x \to x', y) = \left[U_m(x',y)  \,\reallywidetilde{\left(U_m^\dagger(x,y)  D(\alpha) U_m(x,y)\right)} \,U_m^\dagger(x',y)\right]^{\beta}_{\beta} \, ,
\end{equation}
where $D(\alpha)$ is a diagonal matrix with a non-vanishing (unit) component, that is $\left[D(\alpha)\right]^\beta_{\gamma} = \delta_\alpha^\beta \delta^\alpha_\gamma$ (no summation). Equation \eqref{eq:defATAO} corresponds to the probability for a state of flavour $\alpha$ and momentum $y$ generated at a scale factor $x$, ``averaged'' according to the ATAO approximation, to re-emerge as a flavour $\beta$ at later $x'$, if it is not affected by collisions in the meantime.
When evaluated at $x' \to \infty$, the asymptotic $\mathcal{T}(\alpha \to \beta, x, y) \equiv \mathcal{T}(\alpha \to \beta, x \to \infty, y)$ provide information on neutrino flavour conversion from their last scattering with other species, until all neutrino spectra are frozen since mean-field and collisions are then negligible (figure~\ref{fig:ATAO}).

If mean-field effects can be ignored, the asymptotic ATAO transfer function converges to the following expression
\begin{equation}
\label{VacuumAverage}
\mathcal{T}^{\rm vac}(\alpha \to \beta) \equiv \left[U \,\reallywidetilde{\left(U^\dagger D(\alpha) U \right)}\,U^\dagger\right]^{\beta}_{\beta}\,,
\end{equation}
which is independent of $y$ and where the PMNS matter matrix is replaced by the vacuum one.

To gather further insight on the impact of the mixing and mean-field terms, we have performed two schematic calculations, including either the neutrino probabilities at the end of the evolution, i.e. $T_{\mathrm{cm}, f} = 0.01 \, \mathrm{MeV}$ (``NO, post-aver.''), or keeping only the mixing and collision terms during the evolution (``without mean-field''). The corresponding results are shown in table~\ref{Table:Res_NuDec}.

In the first schematic calculation, we have introduced a post-averaging of the no-oscillation results as
\begin{equation}
\label{eq:DefPost}
(\vrho^{\rm post})^\beta_{\beta} = \sum_\alpha (\vrho^{\rm NO})^\alpha_\alpha \, \mathcal{T}^{\rm vac}(\alpha \to \beta)\, .
\end{equation}
From table~\ref{Table:Res_NuDec} one can see that the electronic spectra are suppressed and other neutrino types spectra are enhanced by the vacuum averaging procedure. One can nearly recover the oscillation case results by averaging the final results found without oscillations, thus showing that the different values of the effective neutrino temperatures between the no-oscillation case and the full oscillation case are likely to be due to the effect of the mixings. However, the post averaging of the no-oscillation case (which by construction preserves $\Neff$) does not capture the enhancement of $N_{\rm eff}$ of the full oscillation case, discussed at the end of section~\ref{subsec:results}.

In the second schematic calculation we have solved the QKEs \eqref{eq:QKE_final} without the mean-field term, i.e., keeping only the vacuum and collision terms\footnote{We thus have $U_m = U$ and the matter basis is the mass basis.} (table~\ref{Table:Res_NuDec}). This is somehow an improvement of the ``post averaging'' procedure, since it neglects the variation of the transfer functions (which always have their asymptotic vacuum values), but accounts correctly for the effect of collisions. The accuracy of the results compared to the full treatment shows once more that the effect of the mean-field is very mild in this case. Indeed, the mean-field contribution becomes effective when $\vrho$ deviates from a matrix proportional to the identity, which only happens when $x \sim 3\times10^{-1}$: however at this point the mean-field contribution is becoming negligible compared to the vacuum one (cf.~figure~\ref{fig:ATAO}). Note that this would not hold if we introduced chemical potentials \cite{Damping98,Dolgov_NuPhB2002,Gava:2010kz,Mirizzi2012,Saviano2013,HannestadTamborra}. The higher value obtained for $\Neff$ in this case can be qualitatively understood. Since ${\cal T}^{\rm vac}(e \to e) < {\cal T}(x \ll1, e \to e)$, $\nu_e$ produced by collisions will be more converted into other flavours (in particular $\nu_\tau$) at early times compared to the full calculation. This frees some phase space for the reheating of $\nu_e$, which is the dominant process. More entropy is transferred from $e^\pm$ annihilations, which increases slightly $\Neff$.

These transfer functions also shed some light on the importance of the precise value of the mixing angles, which explain some discrepancy with previous results (see section~\ref{subsec:results}). Indeed, varying $\theta_{ij}$ within their uncertainty ranges slightly modify the ${\cal T}(\alpha \to \beta)$ curves, which can cross each other. For instance, with the set of parameters used in \cite{Relic2016_revisited}, the asymptotic value ${\cal T}^{\rm vac}(e \to \tau)$ is higher than ${\cal T}^{\rm vac}(e \to \mu)$, contrary to figure~\ref{fig:ATAO}. This higher conversion of electron neutrinos into tau neutrinos explains why their final temperatures are $z_{\nu_\tau} \gtrsim z_{\nu_\mu}$ (the values remaining very close).

\paragraph{Sensitivity to the mass hierarchy.} 
In the inverted hierarchy, for which $\Delta m_{31}^2 < 0$, $N_{\rm eff}$ is increased by $5\times10^{-6}$. In this case, $\nu_e$ can be generated above an MSW resonance (e.g.~at about $4\,{\rm MeV}$ for $y=5$), and are converted nearly entirely as $\nu_\mu$ and $\nu_\tau$ (figure~\ref{fig:ATAO}). Again, this impacts subsequent collisions because it frees some phase space for $\nu_e$, which is beneficial for the total production of neutrinos. However, since neutrino decoupling occurs mainly at temperatures which are below the MSW resonance,\footnote{This is not the case for very large $y$ but they are subdominant in the total energy density budget.} the differences between normal and inverted hierarchies are extremely small. 

To summarise, neutrino decoupling is mostly sensitive to the neutrino mixings, whereas it has little sensitivity to the mass-squared differences and therefore to the neutrino mass hierarchy.

\section{Flavour oscillations and Big Bang nucleosynthesis}
\label{sec:BBN}

Predicting a precise value of $\Neff$ in the standard cosmological model is timely since forthcoming generations of CMB experiments aim at measuring a possible contribution of light relic particles predicted by extensions of the standard model \cite{CMB-S4}. 
Yet CMB is not the only cosmological stage impacted by neutrinos, and $\Neff$ can be further constrained using the predicted abundances of light elements produced during BBN.

Indeed, incomplete neutrino decoupling, by giving rise to slightly non-thermal spectral distortions in neutrino spectra and modifying the photon to neutrino temperature ratio, affects BBN in various ways (see ref.~\cite{Pitrou_2018PhysRept} for a review).
\begin{enumerate}
	\item The neutron-to-proton ratio freezes out from equilibrium when the rates of $n\leftrightarrow p$ interconversion reactions ($n + \nu_e \leftrightarrow p + e^-$, $n + e^+ \leftrightarrow p + \bnu_e$, $n \leftrightarrow p + e^- + \bnu_e$) drop below the Hubble expansion rate. The neutron fraction $X_n \equiv n_n/n_b$, with $n_b$ the baryon density, thus depends on when freeze-out occurs, and on deviations from standard nuclear statistical equilibrium that all depend on $z$, $z_{\nu_e}$ and $\delta g_{\nu_e}$ \cite{Froustey2019}.
	\item After the freeze-out, the neutron fraction decreases since neutrons continue to undergo beta decay until the onset of nucleosynthesis at $T_\gamma = \TNuc$. The higher energy density of neutrinos for a given photon temperature, parametrized by $\Neff$, increases the Hubble rate compared to the instantaneous decoupling case, thus diminishing the number of neutrons that decayed. This is the so-called \emph{clock effect} \cite{Dodelson_Turner_PhRvD1992,Fields_PhRvD1993}, which tends to increase the fraction of neutrons at the beginning of nucleosynthesis $X_n(\TNuc)$, and consequently the helium fraction $\YP \equiv 4 n_{\He{4}}/n_b \simeq 2 X_n(\TNuc)$ since these neutrons are almost fully converted into $\He{4}$.
	\item The production of other light elements from the remaining traces of neutrons is also controlled by the clock effect \cite{Grohs2015,Froustey2019}. For instance, deuterium is mainly destroyed from its equilibrium value to its frozen-out abundance at the end of BBN \cite{SmithBBN}. The higher expansion rate leaves less time for this destruction to happen, which leads to a net increase of the deuterium abundance.
\end{enumerate}
There was some discrepancy in the literature about the sign of variation of the different abundances due to these effects, see for instance table 3 in \cite{Mangano2005} and table V in \cite{Grohs2015}. The extensive analysis of ref.~\cite{Froustey2019} favoured the latter results. Though it did not include flavour oscillations, it predicted that the main conclusions would hold since the final neutrino spectra are qualitatively similar to the no-oscillations case, only ``averaged''. 

We aim at filling this gap and therefore introduce the results from section \ref{subsec:results} in the BBN code \texttt{PRIMAT} \cite{Pitrou_2018PhysRept}. This section is meant as an extension of the work \cite{Froustey2019}, from which we will borrow the notation. We implement neutrino-induced corrections following the three levels of refinements introduced in \cite{Froustey2019} $i$) assuming that the three neutrino species have thermal spectra at the average temperature $\widehat{T}_\nu \equiv \frac13 (T_{\nu_e}^4 + T_{\nu_\mu}^4 + T_{\nu_\tau}^4)^{1/4}$ (``$\widehat{T}_\nu$''), $ii$) using the proper effective temperature for $\nu_e$, but without non-thermal distortions (``$T_{\nu_e}$, no distortions''), and $iii$) using the real spectra from \texttt{NEVO} (``$T_{\nu_e}$, with distortions''). Note that the total neutrino energy density, so $\Neff$, is identical in all three implementations, therefore the clock effect contributions will be identical. We report the obtained values for the abundances of helium-4, deuterium, helium-3 and lithium-7 in table~\ref{Table:Full_Corrections}, with the associated relative variations compared to the instantaneous decoupling case\footnote{The instantaneous decoupling baseline is the same with or without flavour oscillations, since in this limit all three neutrino species have FD spectra at the comoving temperature $\Tcm$.} in table~\ref{Table:Full_Delta}.

Note that a few updates were made to \texttt{PRIMAT} compared to previous implementations~\cite{Pitrou_2018PhysRept,Froustey2019}: we used the latest values of the physical constants and cosmological parameters such as the neutron lifetime ($\tau_n = 879.4 \, \mathrm{s}$), the axial coupling of nucleons ($g_A = 1.2756$)~\cite{PDG} or the baryon density ($\Omega_b h^2 = 0.0224$)~\cite{Planck18}, and included QED corrections due to electron-positron pair production to some nuclear rates~\cite{PitrouPospelov20}.

\renewcommand{\arraystretch}{1.4}

\begin{table}[!ht]
	\centering
	\begin{tabular}{|l|cccc|}
  	\hline 
 BBN framework & $\YP$  & ${\rm  D}/{\rm H}  \times 10^5$ &  ${\He3}/{\rm  H} \times 10^5$ &  ${\Li}/{\rm  H} \times 10^{10}$ \\
  \hline \hline
  Inst. decoupling & $0.24711$  & $2.4167$ & $1.0690$  & $5.8006$ \\ \hline
    $\widehat{T}_\nu$ (NO) & $0.24716$  & $2.4256$ & $1.0703$  & $5.7768$ \\
    $T_{\nu_e},$ with distortions (NO) & $0.24716$  & $2.4256$ & $1.0703$  & $5.7767$ \\ \hline
  $\widehat{T}_\nu$ & $0.24716$  & $2.4258$ & $1.0703$  & $5.7764$ \\
    $T_{\nu_e},$ no distortions & $0.24713$  & $2.4256$  & $1.0703$  & $5.7759$ \\
  $T_{\nu_e},$ with distortions & $0.24721$  & $2.4261$ & $1.0703$  & $5.7772$ \\ \hline
\end{tabular}
	\caption{Light element abundances, including all weak rate corrections \cite{Pitrou_2018PhysRept} and QED corrections up to ${\cal O}(e^3)$ to plasma thermodynamics, for various implementations of neutrino-induced corrections. $\He3$ stands for $(\He3 + \mathrm{T})$ and $\Li$ stands for $(\Li + \Be)$ to account for slow radioactive decays. We compare the inclusion of results from neutrino decoupling with and without (NO) oscillations.
	\label{Table:Full_Corrections}}
\end{table}

\begin{table}[!ht]
	\centering
	\begin{tabular}{|l|cccc|}
  	\hline 
 BBN framework & $\delta \YP \, (\%)$  & $\delta ({\rm  D}/{\rm H}) \, (\%)  $ &  $\delta ({\He3}/{\rm  H}) \, (\%) $ &  $\delta ({\Li}/{\rm  H}) \, (\%) $ \\
  \hline \hline
    $\widehat{T}_\nu$ (NO) & $0.020$  & $0.369$ & $0.120$  & $-0.411$ \\
    $T_{\nu_e},$ with distortions (NO) & $0.020$  & $0.368$ & $0.120$  & $- 0.412$ \\ \hline
  $\widehat{T}_\nu$ & $0.021$  & $0.375$ & $0.122$  & $-0.418$ \\
    $T_{\nu_e},$ no distortions & $0.007$  & $0.367$  & $0.120$  & $-0.427$ \\
  $T_{\nu_e},$ with distortions & $0.042$  & $0.387$ & $0.126$  & $- 0.404$ \\ \hline
\end{tabular}
	\caption{Relative variations of the light element abundances compared to the instantaneous decoupling limit, in the same frameworks as table~\ref{Table:Full_Corrections}.
	\label{Table:Full_Delta}}
\end{table}

The variation of the $\He4$ abundance due to incomplete neutrino decoupling is estimated by
\begin{equation}
\delta \YP = \delta X_n^{\rm [Nuc]} = \delta X_n^{\rm [FO]} + \delta X_n^{[\Delta t]} \, ,
\end{equation}
where the first equality comes from the almost total conversion of free neutrons into $\He{4}$. $\delta X_n^{\rm [FO]}$ is the variation of the neutron fraction at freeze-out (point 1 above), and $\delta X_n^{[\Delta t]}$ is the variation due to the different duration of beta decay (clock effect, point 2 above). The variation of the other abundances relative to the proton fraction $i/{\rm H} \equiv n_i/n_{\rm H}$ is given by \cite{Froustey2019}
\begin{equation}
\delta (i / {\rm H}) \simeq \delta X_i^{[\Delta t]} + \delta \YP \, ,
\end{equation}
where $\delta X_i^{[\Delta t]}$ is the variation of the final abundance due to the clock effect, and the $\delta \YP$ contribution is actually $- \delta X_{\rm H}$.

\paragraph{Comparison of implementations.}

\begin{figure}[!ht]
	\centering
	\includegraphics[width=10cm]{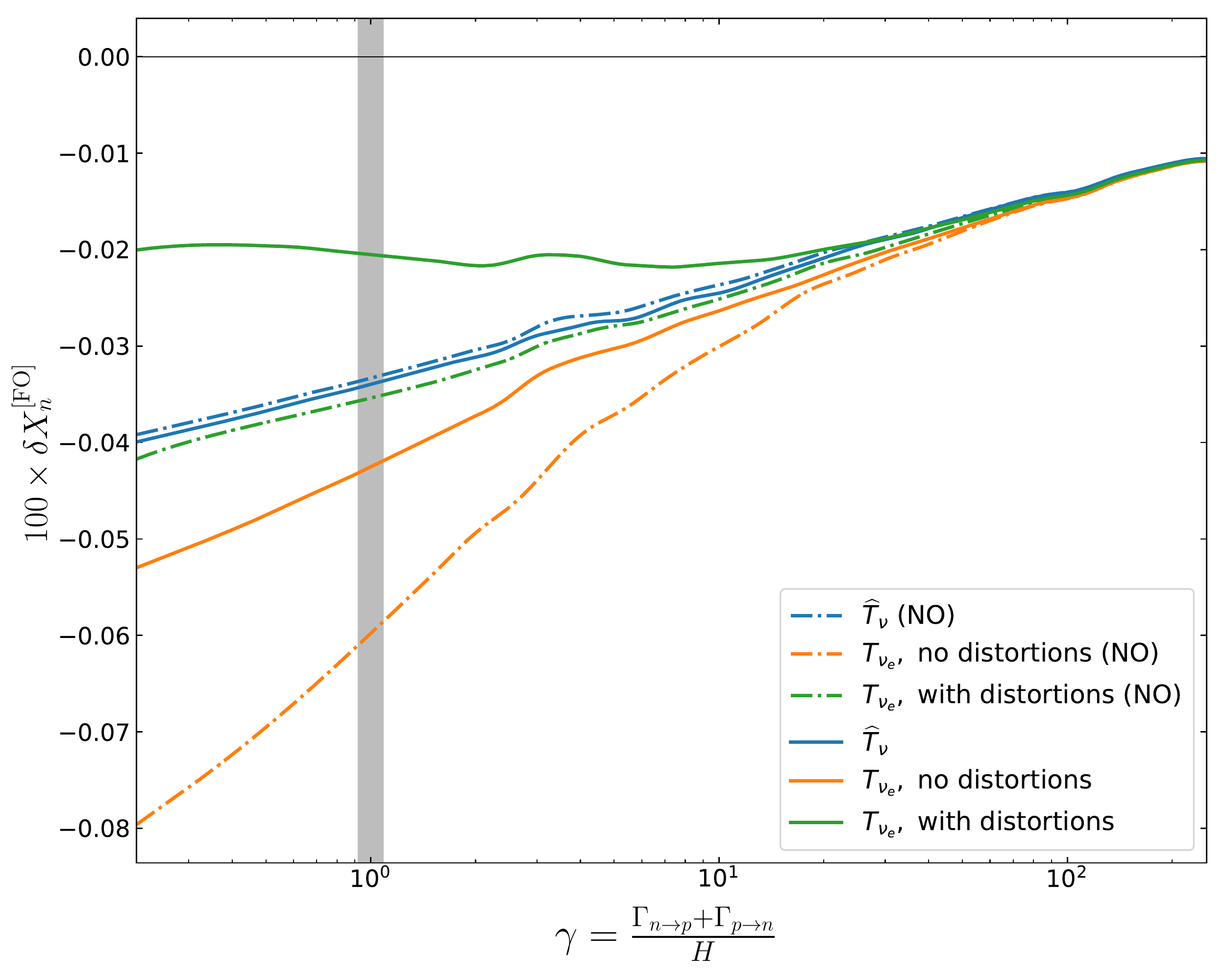}
	\caption{\label{fig:deltaXn} Neutron fraction variation around freeze-out, for different implementations of neutrino-induced corrections, with and without flavour oscillations.}
\end{figure}

An a priori surprising conclusion of ref.~\cite{Froustey2019} was the quasi-equivalence of the ``$\widehat{T}_\nu$'' and full implementations. We recover this feature in the no-oscillation case (cf.~for instance the first two lines of table~\ref{Table:Full_Delta}), while there is a sizeable difference when using the neutrino spectra with oscillations. Let us focus on the helium fraction $\YP$. We plot the variation of the neutron fraction at freeze-out $\delta X_n^{\rm [FO]}$ on figure~\ref{fig:deltaXn}. First note that the average temperature implementations give quasi identical results with and without oscillations, which is a direct consequence of the small difference of $\Neff$ in table~\ref{Table:Res_NuDec}. Then, including the true $T_{\nu_e}$ reduces $\delta X_n^{\rm [FO]}$: since $T_{\nu_e} > \widehat{T}_\nu$, the weak rates increase and freeze-out is delayed, thus $X_n$ tracks its equilibrium value longer. This reduction of $X_n^{\rm [FO]}$ is more important without oscillations because the effective $\nu_e$ temperature is much higher than the average temperature in this case (cf.~figure~\ref{fig:Tnu}). Finally, the spectral distortions alter the detailed balance relation which sets the neutron-to-proton ratio \cite{Froustey2019}, shifting $X_n^{\rm [FO]}$ in the opposite direction. Once again, this re-increase of the neutron fraction is more important in the no-oscillation case, since $\lvert \delta g_{\nu_e}^{\rm (NO)} \rvert >\lvert \delta g_{\nu_e} \rvert$ (cf.~figure~\ref{fig:deltagnu}).

All in all, the final value of $\delta X_n^{\rm [FO]}$ is higher with oscillations, and exceeds the average temperature value (i.e., the solid green curve is above the solid blue one, while the dash-dotted green and blue curves almost coincide). This could be surprising, since $z_{\nu_e}$ and $\delta g_{\nu_e}$ are both reduced by about 25 \% with mixing (figures~\ref{fig:Tnu} and \ref{fig:deltagnu}), so we would expect the solid curves to be in homothetic ratio with the dash-dotted ones. However, $z_{\nu_e}$ is reduced by 25 \% \emph{compared to $z=1$}, but is much more reduced, by $\sim $ 68 \% \emph{compared to $\hat{z}$}. That is why the gap between the solid blue and orange curves is 68 \% smaller than the gap between the dash-dotted blue and orange curves. Since the up-shifting of $X_n$ due to distortions is just reduced by $\sim$ 10 \%,\footnote{It is not a 25 \% reduction since the relation between $\delta g_{\nu_e}$ and the modification of detailed balance is not exactly linear.} the ``$T_{\nu_e}$, with distortions'' value in the oscillation case is higher.

Note however that, although the average temperature implementation is less accurate in the oscillation case than in the no-oscillation case, it is sufficient to provide the various abundances at a relative precision of a few $10^{-4}$, which is well beyond experimental uncertainties. Moreover, this method is particularly simple, since all information is contained in one parameter, the average effective temperature $\widehat{T}_{\nu}(T_\gamma)$. It can be used in a BBN code from a table of its values, or be deduced from the dimensionless heating function $\mathcal{N}$ \cite{Parthenope,Parthenope_reloaded} that parametrizes the heat transfer from $e^\pm$ annihilations and which can be fitted to the desired precision.

$\Neff$ having the same value in all three implementations, the difference between the last three lines of tables~\ref{Table:Full_Corrections} and \ref{Table:Full_Delta} lies in the variation of $\delta X_n^{\rm [FO]}$. This is somehow hidden for $\He3$ and $\Li$ because they are the aggregated results of $(\He3 + {\rm T})$ and $(\Li + \Be)$ respectively.

\paragraph{Overall effect of flavour oscillations.} Let us now discuss the global difference in the final abundances due to the inclusion of oscillations. To keep the discussion simple, we will discuss the average temperature implementation, where all the information about neutrino spectra is encoded in $\widehat{T}_\nu$ (the differences between the three implementations for a given \mbox{(no-)oscillation} case being explained above). We see from table~\ref{Table:Res_NuDec} that $\Neff$ is slightly higher when including oscillations, thus increasing the clock effect. For instance, there will be less time for the destruction of deuterium to take place, and we expect a higher ${\rm D}$ abundance. The same argument goes for $\He3$ and $\rm T$, causing an increase of $\He3/{\rm H}$. Last, the abundance $\Li/{\rm H}$ is dominated by primary $\Be$, that is produced during nucleosynthesis: a faster expansion diminishes the $\Be$ yield, and thus the value of $\Li/{\rm H}$. The results corresponding to the cases $\widehat{T}_\nu$ (NO) and $\widehat{T}_\nu$ in tables~\ref{Table:Full_Corrections} and \ref{Table:Full_Delta} can be understood using these simple heuristic arguments.

\section{Conclusions}

We derived the QKEs governing neutrino evolution at the epoch of weak decoupling using a BBGKY-like formalism, obtaining the mean-field terms up to ${\cal O}(1/m_{W,Z}^2)$ order and the collision terms with their full matrix structure. We solved the QKEs and presented the ATAO approximation which allows to increase the computation speed. This approximation is based on the assumptions that there is a clear separation of time-scales between the oscillation frequencies and the collision rate, the off-diagonal terms of the density matrix in the matter basis are averaged out and the matter basis evolves adiabatically.

Results on $N_{\rm eff}$ and the neutrino final spectra were presented with a numerical precision better than $10^{-4}$. A better precision would require the inclusion of several corrections. First, one would need to consider QED effects in the collision rates \cite{Passera_QED,Esposito_QED}, and further corrections to the plasma thermodynamics at order $e^4$ and sub-leading logarithmic-dependent terms at order $e^2$~\cite{Bennett2020}. But more importantly it would not be possible to consider a homogeneous cosmology since fluctuations inherited from the inflationary phase, and imprinted in the CMB, are of order $10^{-5}$. One would then need to consider fluctuations in the QKE as was done to estimate fluctuations in the CMB.\footnote{Furthermore, the physics of decoupling also depends on the Fermi and Newton constants, and the latter is only known with a $4 \times 10^{-5}$ precision.} 

The obtained value of $\Neff \simeq 3.0440$ and the associated spectral distortions were used in the BBN code \texttt{PRIMAT} to investigate the consequences of incomplete neutrino decoupling \emph{with} flavour oscillations on the primordial production of light elements, solving the discrepancy between \cite{Mangano2005} and \cite{Grohs2015}, with results in agreement with \cite{Froustey2019}. Even though the subsequent variations occur at precisions well beyond experimental uncertainties, we were able to understand the physical processes at play, thus checking the validity of our results. The nuclear abundances, with all weak rates corrections included (as in \cite{Pitrou_2018PhysRept}), and taking completely into account neutrino distorted spectra, are reported in the last line of table~\ref{Table:Full_Corrections}. The next update of \texttt{PRIMAT} will include these results.

\acknowledgments

The authors would like to thank Evan Grohs for useful discussions.

\appendix

\section{Derivation of the formal collision term}
\label{app:derivation_collision}

Compared to the Boltzmann treatment of neutrino evolution, which neglects flavour mixing, the QKE contains mean-field terms, and the collision term has a richer matrix structure with non-zero off-diagonal components. To derive this collision term, i.e., the contribution to the evolution of the one-body density matrix from two-body correlations, one needs an expression for the correlated part $C$ in \eqref{eq:eqvrho}. It is obtained from the evolution equation for $\vrho^{(12)}$, where we separate correlated and uncorrelated parts \cite{Lac04}.

To do so, we need a splitting similar to \eqref{eq:splitrho} for the three-body density matrix,
\begin{equation}
\label{eq:splitrho123}
\vrho^{ikm}_{jln} = 6 \vrho^{i}_{[j}\vrho^{k}_{l}\vrho^{m}_{n]} + 9 \vrho^{[i}_{[j} C^{km]}_{ln]} + C^{ikm}_{jln} \, .
\end{equation}
This allows \eqref{eq:hierarchy} to be rewritten as an equation for the two-body correlation function \cite{Volpe_2013}. In the \emph{molecular chaos} ansatz, correlations are built through a collision between uncorrelated particles. These correlations then evolve ``freely'', i.e., we do not take into account a mean-field background for $C$. The evolution equation is thus greatly simplified, retaining only the vacuum and Born terms:
\begin{equation}
\begin{aligned}
i  \frac{\dd C^{ik}_{jl}}{\dd t} &= \left[t^{i}_{r} C^{rk}_{jl} + t^{k}_{p} C^{ip}_{jl} - C^{ik}_{rl} t^{r}_{j} - C^{ik}_{jp} t^{p}_{l} \right] \\
&\quad + (\hat{1} - \vrho)^i_r (\hat{1}-\vrho)^k_p \, \tilde{v}^{rp}_{sq} \, \vrho^{s}_{j} \vrho^{q}_{l} -  \vrho^i_r \vrho^k_p \, \tilde{v}^{rp}_{sq} \, (\hat{1} -\vrho)^{s}_{j} (\hat{1} - \vrho)^{q}_{l} \, ,
\end{aligned}
\end{equation}
where the second line will be labelled $B^{ik}_{jl}$. We can actually solve this equation, starting from $C(t=0)=0$,
\begin{equation}
\label{eq:solveC}
C^{ik}_{jl}(t) = - i \int_{0}^{t}{\dd s \,  T^{ik}_{mp}(t,s) B^{mp}_{nq}(s) {T^\dagger}^{nq}_{jl}(t,s)} \, ,
\end{equation}
with the evolution operator
\begin{equation}
T^{ik}_{jl}(s,s') = \exp{\left(-i \int_{s'}^{s}{\dd \tau \, \hat{t}(\tau)}\right)}^{i}_{j} \exp{\left(-i \int_{s'}^{s}{\dd \tau \, \hat{t}(\tau)}\right)}^{k}_{l} \, .
\end{equation}
Now we consider that there is a clear separation of scales \cite{SiglRaffelt}, hence the duration of one collision is very small compared to the variation timescale of the density matrices (i.e., compared to the duration between two collisions, and the typical inverse oscillation frequency). Therefore, the argument inside the integral of \eqref{eq:solveC} is only non-zero for $s \simeq 0$: we can extend the integration domain to $+ \infty$, while the operators keep their $t=0$ value. Finally we symmetrize the integration domain\footnote{See section 6.1 in ref.~\cite{FidlerPitrou} for a detailed discussion of this procedure.} with respect to 0 (with an extra factor of $1/2$), which leads to the equation with collision term:
\begin{align}
i  \frac{\dd \vrho^{i}_{j}}{\dd t} &= \left[ \hat{t} + \hat{\Gamma}, \hat{\varrho}\right]^{i}_{j} - \frac{i}{4} \int_{-\infty}^{+\infty}{\dd t\,  \left[ \tilde{v},T(t,0)B(0) T^\dagger(t,0) \right]}^{ik}_{jk} \\ 
&= [(t^i_k + \Gamma^i_k)\varrho^k_j - \varrho^i_k (t^k_j + \Gamma^k_j)] \nonumber \\
&\qquad  - \frac{i}{4}  \underbrace{\int_{-\infty}^{+\infty}{\dd t \, e^{-i(E_m+E_l-E_j-E_k)t}}}_{(2\pi) \delta(E_m + E_l - E_j - E_k)}\left[\tilde{v}^{ik}_{rl} B^{rl}_{jk} - B^{ik}_{rl} \tilde{v}^{rl}_{jk} \right] \, , \\
&\equiv \left[ \hat{t} + \hat{\Gamma}, \hat{\varrho}\right]^{i}_{j} + i \, \hat{\mathcal{C}}^{i}_{j}
\end{align}
The exponential of energies comes from the $T$ terms, using that the density matrix for a given momentum $\vrho(p)$ satisfies $\hat{t}{\vrho}(p) = p \, {\varrho}(p)$.

\section{Interaction potential matrix elements}
\label{app:matrixelements}

The relevant two-body interactions correspond to standard model interactions involving neutrinos and antineutrinos. In the early universe, they interact throught weak processes with electrons, positrons and other (anti)neutrinos. Therefore, we must take as interaction Hamiltonian \eqref{eq:defHint} the useful part of the standard model Hamiltonian of weak interactions, that is given by
\begin{equation}
\label{eq:defHsm}
\hat{H}_{\rm int} = \hat{H}_{CC} + \hat{H}_{NC}^{\rm mat} + \hat{H}_{NC}^{\nu \nu} \, ,
\end{equation}
where we separated three contributions:
\begin{itemize}
	\item the charged current hamiltonian,
\begin{multline}
\label{eq:hcc}
\hat{H}_{CC} = 2 \sqrt{2} G_F m_W^2 \int{\ddp{1} \ddp{2} \ddp{3} \ddp{4}} \ (2\pi)^3 \delta^{(3)}(\vp_1 + \vp_2 - \vp_3 - \vp_4) \\ \times  [\bar{\psi}_{\nu_e}(\vp_1)\gamma_\mu P_L\psi_e(\vp_4)] W^{\mu \nu}(\Delta) [\bar{\psi}_e(\vp_2) \gamma_\nu P_L \psi_{\nu_e}(\vp_3)] \, ,
\end{multline}
with $\psi(\vec{p}) = \sum_{h} \left[ \ha(\vec{p},h) u^h(\vec{p})+ \hbd(-\vec{p},h) v^h(-\vec{p}) \right]$ the Fourier transform of the quantum fields, $P_L = (1-\gamma_5)/2$ the left-handed projection operator, and the gauge boson propagator
\begin{equation}
\label{eq:propagator}
W^{\mu \nu}(\Delta) = \frac{g^{\mu \nu} - \frac{\Delta^\mu \Delta^\nu}{m_W^2}}{m_W^2 - \Delta^2} \simeq \frac{g^{\mu \nu}}{m_W^2} + \frac{1}{m_W^2}\left(\frac{\Delta^2 g^{\mu \nu}}{m_W^2} - \frac{\Delta^\mu \Delta^\nu}{m_W^2}\right) \, .
\end{equation}
The lowest order in this expansion is the usual 4-Fermi effective theory. The momentum transfer is $\Delta = p_1-p_4$ for a $t$-channel ($\nu_e-e^-$ scattering), and $\Delta= p_1 + p_2$ for the $s$-channel ($\nu_e-e^+$). At Fermi order, we get for instance (after a Fierz transformation):
\begin{multline}
\tilde{v}^{\nu_e(1) e(2)}_{\nu_e(3) e(4)} = 2 \sqrt{2} G_F \, (2 \pi)^3 \delta^{(3)}(\vec{p}_1 + \vec{p}_2 - \vec{p}_3 - \vec{p}_4) \\
\times [\bar{u}_{\nu_e}^{h_1} (\vec{p}_1) \gamma^\mu P_L u_{\nu_e}^{h_3} (\vec{p}_3)] \ [\bar{u}_{e}^{h_2} (\vec{p}_2)\gamma_\mu P_L u_{e}^{h_4} (\vec{p}_4)] \, .
\end{multline}

	\item the neutral current interactions with the matter background (electrons and positrons),
\begin{multline}
\label{eq:hncmat}
\hat{H}_{NC}^{\rm mat} = 2 \sqrt{2} G_F m_Z^2 \sum_{\alpha} \int{\ddp{1} \ddp{2} \ddp{3} \ddp{4}} \ (2\pi)^3 \delta^{(3)}(\vp_1 + \vp_2 - \vp_3 - \vp_4) \\ \times  [\bar{\psi}_{\nu_\alpha}(\vp_1)\gamma_\mu P_L\psi_{\nu_\alpha}(\vp_3)] Z^{\mu \nu}(\Delta) [\bar{\psi}_e(\vp_2) \gamma_\nu (g_L P_L + g_R P_R) \psi_e(\vp_4)] \, ,
\end{multline}
where $Z^{\mu \nu}$ is identical to $W^{\mu \nu}$ with the replacement $m_W \to m_Z$. The neutral-current couplings are $g_L = -1/2 + \sin^2{\theta_W}$ and $g_R = \sin^2{\theta_W}$, where $\sin^2{\theta_W} \simeq 0.231$ is the weak-mixing angle.
	\item the self-interactions of neutrinos,\footnote{To understand the different prefactor from $\hat{H}_{NC}^{\rm mat}$, start from the general neutral-current Hamiltonian:
\[
\hat{H}_{NC} = 2 \sqrt{2} G_F m_Z^2 \sum_{f,f'} \int{\cdots \ \left[\bar{\psi}_{f} \gamma_\mu (g_L^f P_L + g_R^f P_R) \psi_f\right]Z^{\mu \nu}(\Delta)\left[\bar{\psi}_{f'}\gamma_\nu (g_L^{f'} P_L + g_R^{f'} P_R) \psi_{f'}\right]} \]
Now the multiplicity of each term and the use of $g_L^\nu = 1/2$, $g_R^\nu=0$ lead to the Hamiltonians above.
}
\begin{multline}
\label{eq:hncself}
\hat{H}_{NC}^{\nu \nu} = \frac{G_F}{\sqrt{2}} m_Z^2 \sum_{\alpha, \beta} \int{\ddp{1} \ddp{2} \ddp{3} \ddp{4}} \ (2\pi)^3 \delta^{(3)}(\vp_1 + \vp_2 - \vp_3 - \vp_4) \\ \times  [\bar{\psi}_{\nu_\alpha}(\vp_1)\gamma_\mu P_L\psi_{\nu_\alpha}(\vp_3)] Z^{\mu \nu}(\Delta) [\bar{\psi}_{\nu_\beta}(\vp_2) \gamma_\nu P_L \psi_{\nu_\beta}(\vp_4)] \, .
\end{multline}
\end{itemize}

We show in table~\ref{Table:MatrixElements} the set of interaction matrix elements derived from these Hamiltonians, which are needed for the neutrino collision term. To compute the mean-field potentials at order $1/m_{W,Z}^2$, one needs the matrix elements from the expansion of the propagator \eqref{eq:propagator}, which are obtained similarly and not reproduced here for the sake of brevity.

\renewcommand{\arraystretch}{1.3}

\begin{table}[!h]
	\centering
	\begin{tabular}{|l|r|}
  	\hline 
 Interaction process &  $\tilde{v}^{12}_{34}/\left[\sqrt{2} G_F (2 \pi)^3 \delta^{(3)}(\vp_1+\vp_2-\vp_3-\vp_4)\right]$   \\
  \hline \hline
  $CC$ &   \\ \hline
  $\nu_e(1) e(2) \nu_e(3) e(4)$ & $2 \times [\bar{u}_{\nu_e}^{h_1} (\vec{p}_1) \gamma^\mu P_L u_{\nu_e}^{h_3} (\vec{p}_3)]  [\bar{u}_{e}^{h_2} (\vec{p}_2)\gamma_\mu P_L u_{e}^{h_4} (\vec{p}_4)]$  \\
    $\nu_e(1) \bar{e}(2) \nu_e(3) \bar{e}(4)$ & $- 2 \times [\bar{u}_{\nu_e}^{h_1} (\vec{p}_1) \gamma^\mu P_L u_{\nu_e}^{h_3} (\vec{p}_3)] [\bar{v}_{e}^{h_4} (\vec{p}_4)\gamma_\mu P_L v_{e}^{h_2} (\vec{p}_2)]$  \\
  $\nu_e(1) \bar{\nu}_e(2) e(3) \bar{e}(4)$ & $2 \times [\bar{u}_{\nu_e}^{h_1} (\vec{p}_1) \gamma^\mu P_L v_{\nu_e}^{h_2} (\vec{p}_2)]  [\bar{v}_{e}^{h_4} (\vec{p}_4)\gamma_\mu P_L u_{e}^{h_3} (\vec{p}_3)]$  \\ \hline \hline
  $NC, \text{matter}$ &   \\ \hline
  $\nu_e(1) e(2) \nu_e(3) e(4)$ & $2 \times [\bar{u}_{\nu_e}^{h_1} (\vec{p}_1) \gamma^\mu P_L u_{\nu_e}^{h_3} (\vec{p}_3)]  [\bar{u}_{e}^{h_2} (\vec{p}_2)\gamma_\mu (g_L P_L + g_R P_R) u_{e}^{h_4} (\vec{p}_4)]$  \\
    $\nu_e(1) \bar{e}(2) \nu_e(3) \bar{e}(4)$ & $- 2 \times [\bar{u}_{\nu_e}^{h_1} (\vec{p}_1) \gamma^\mu P_L u_{\nu_e}^{h_3} (\vec{p}_3)] [\bar{v}_{e}^{h_4} (\vec{p}_4)\gamma_\mu (g_L P_L + g_R P_R) v_{e}^{h_2} (\vec{p}_2)]$  \\
  $\nu_e(1) \bar{\nu}_e(2) e(3) \bar{e}(4)$ & $2 \times [\bar{u}_{\nu_e}^{h_1} (\vec{p}_1) \gamma^\mu P_L v_{\nu_e}^{h_2} (\vec{p}_2)]  [\bar{v}_{e}^{h_4} (\vec{p}_4)\gamma_\mu (g_L P_L + g_R P_R) u_{e}^{h_3} (\vec{p}_3)]$  \\ \hline  \hline
    $NC, \text{self-interactions}$ &   \\ \hline
  $\nu_\alpha(1) \nu_\beta(2) \nu_\alpha(3) \nu_\beta(4)$ & $(1+\delta_{\alpha \beta})
\times [\bar{u}_{\nu_\alpha}^{h_1} (\vec{p}_1) \gamma^\mu P_L u_{\nu_\alpha}^{h_3} (\vec{p}_3)]  [\bar{u}_{\nu_\beta}^{h_2} (\vec{p}_2)\gamma_\mu P_L u_{\nu_\beta}^{h_4} (\vec{p}_4)]$  \\
    $\nu_\alpha(1) \bnu_\beta(2) \nu_\alpha(3) \bnu_\beta(4)$ & $- (1 + \delta_{\alpha \beta}) \times [\bar{u}_{\nu_\alpha}^{h_1} (\vec{p}_1) \gamma^\mu P_L u_{\nu_\alpha}^{h_3} (\vec{p}_3)]  [\bar{v}_{\nu_\beta}^{h_4} (\vec{p}_4) \gamma_\mu P_L v_{\nu_\beta}^{h_2} (\vec{p}_2)]$  \\
  $\nu_\alpha(1) \bnu_\alpha(2) \nu_\beta(3) \bnu_\beta(4)$ & $(1+ \delta_{\alpha \beta}) \times [\bar{u}_{\nu_\alpha}^{h_1} (\vec{p}_1) \gamma^\mu P_L v_{\nu_\alpha}^{h_2} (\vec{p}_2)]   [\bar{v}_{\nu_\beta}^{h_4} (\vec{p}_4) \gamma_\mu P_L u_{\nu_\beta}^{h_3} (\vec{p}_3)]$  \\ \hline 
\end{tabular}
	\caption{Interaction matrix elements at lowest order in the expansion of the gauge boson propagators (Fermi effective theory of weak interactions). 
	\label{Table:MatrixElements}}
\end{table}

At leading order, the charged-current processes are written as neutral-current ones thanks to Fierz rearrangement identities. Therefore one can write the global expression:
\begin{multline}
\tilde{v}^{\nu_\alpha(1) e(2)}_{\nu_\beta(3) e(4)} = 2 \sqrt{2} G_F \, (2 \pi)^3 \delta^{(3)}(\vec{p}_1 + \vec{p}_2 - \vec{p}_3 - \vec{p}_4) \\
\times [\bar{u}_{\nu_\alpha}^{h_1} (\vec{p}_1) \gamma^\mu P_L u_{\nu_\beta}^{h_3} (\vec{p}_3)] \ [\bar{u}_{e}^{h_2} (\vec{p}_2)\gamma_\mu (G_L^{\alpha \beta} P_L +  G_R^{\alpha \beta} P_R ) u_{e}^{h_4} (\vec{p}_4)] \, ,
\end{multline}
with, in the Standard model,
\begin{equation}
\label{eq:couplingmatrices}
G_L = \mathrm{diag}(g_L+1,g_L,g_L) \quad , \quad G_R = \mathrm{diag}(g_R,g_R,g_R) \, .
\end{equation}
One can also introduce non-standard interactions which promote the couplings to non-diagonal matrices \cite{Relic2016_revisited}.

\section{Neutrino self-interactions collision term}
\label{app:collision}

As an illustration of the use of the BBGKY formalism to derive the collision integrals, we detail the steps to obtain the neutrino-neutrino scattering contribution to \eqref{eq:C_nn}.

Neutrino-neutrino scattering processes correspond to the terms in \eqref{eq:C11} for which the inner matrix elements are scattering ones $\tilde{v}^{\nu_\delta \nu_\sigma}_{\nu_\delta \nu_\sigma}$. For simplicity, we focus here on the first term in the expression of $\mathcal{C}^{i_1}_{i_1'}$ \eqref{eq:C11}. Here, the index $i_1$ will refer to $\nu_\alpha(\vp_1)$ and $i_1'$ to $\nu_\beta(\vp_{\underline{1}})$. There are two possible contributions to this collision matrix (note that we impose $\vp_k = \vpp_k$ for all $k$, which is enforced by the assumption of homogeneity \eqref{eq:homogeneity}):
\begin{itemize}
	\item when 1 and 3 have the same flavour, the scattering amplitude is:
\begin{align*}
&\tilde{v}^{\nu_\alpha(1) \nu_\gamma(2)}_{\nu_\alpha(3) \nu_\gamma(4)}\times \tilde{v}^{\nu_\delta(3') \nu_\sigma(4')}_{\nu_\delta(1') \nu_\sigma(2')} \\ 
&\quad = 2 G_F^2 \times (2 \pi)^6 \delta^{(3)}(\vp_1 + \vp_2 - \vp_3 - \vp_4) \delta^{(3)}(\vp_1 - \vp_{\underline{1}}) \\
&\quad \quad \times [\bar{u}_{\nu_\alpha}(1) \gamma^\mu P_L u_{\nu_\alpha}(3)][\bar{u}_{\nu_\delta}(3) \gamma^\nu P_L u_{\nu_\delta}(1)] \times  [\bar{u}_{\nu_\gamma}(2) \gamma_\mu P_L u_{\nu_\gamma}(4)][\bar{u}_{\nu_\sigma}(4) \gamma_\nu P_L u_{\nu_\sigma}(2)] \\ 
&\quad = 2 G_F^2  \times (2 \pi)^6 \delta^{(3)}(\vp_1 + \vp_2 - \vp_3 - \vp_4) \delta^{(3)}(\vp_1 - \vp_{\underline{1}}) \\
&\quad \quad \times p_{3 \eta} p_{1 \rho} \tr[\gamma^\rho \gamma^\mu P_L \gamma^\eta \gamma^\nu P_L] \times p_4^\lambda p_2^\tau \tr[\gamma_\tau \gamma_\mu P_L \gamma_\lambda \gamma_\nu P_L] \\
&\quad = 2^5 G_F^2  \times (2 \pi)^6 \delta^{(3)}(\vp_1 + \vp_2 - \vp_3 - \vp_4) \delta^{(3)}(\vp_1 - \vp_{\underline{1}}) \times (p_1 \cdot p_2) (p_3 \cdot p_4)
\end{align*}
With this term, we have the matrix product 
\[\varrho^{\alpha(3)}_{\delta(3)} \varrho^{\gamma(4)}_{\sigma(4)} (1- \varrho)^{\delta(1)}_{\beta(1)} (1- \vrho)^{\sigma(2)}_{\gamma(2)} = \Big[ \Tr[ \vrho_4 \cdot (1-\vrho_2) ] \cdot \vrho_3 \cdot  (1-\vrho_1) \Big]^\alpha_\beta \, . \]
	\item when 1 and 4 have the same flavour, the scattering amplitude is:
\begin{align*}
&\tilde{v}^{\nu_\alpha(1) \nu_\gamma(2)}_{\nu_\gamma(3) \nu_\alpha(4)} \times \tilde{v}^{\nu_\delta(3') \nu_\sigma(4')}_{\nu_\delta(1') \nu_\sigma(2')} \\ &\quad = - 2 G_F^2 \times  (2 \pi)^6 \delta^{(3)}(\vp_1 + \vp_2 - \vp_3 - \vp_4) \delta^{(3)}(\vp_1 - \vp_{\underline{1}}) \\
&\quad \quad \times [\bar{u}_{\nu_\alpha}(1) \gamma^\mu P_L u_{\nu_\alpha}(4)][\bar{u}_{\nu_\sigma}(4) \gamma^\nu P_L u_{\nu_\sigma}(2)] [\bar{u}_{\nu_\gamma}(2) \gamma_\mu P_L u_{\nu_\gamma}(3)][\bar{u}_{\nu_\delta}(3) \gamma_\nu P_L u_{\nu_\delta}(1)] \\ 
&\quad = - 2 G_F^2  \times (2 \pi)^6 \delta^{(3)}(\vp_1 + \vp_2 - \vp_3 - \vp_4) \delta^{(3)}(\vp_1 - \vp_{\underline{1}}) \\
&\quad \quad  \times p_{3 \lambda} p_{1 \rho} p_{4 \eta} p_{2 \tau} \tr[\gamma^\mu P_L \gamma^\eta \gamma^\nu P_L \gamma^\tau \gamma_\mu P_L \gamma^\lambda \gamma_\nu P_L \gamma^\rho] \\
&\quad = 2^5 G_F^2  \times (2 \pi)^6 \delta^{(3)}(\vp_1 + \vp_2 - \vp_3 - \vp_4) \delta^{(3)}(\vp_1 - \vp_{\underline{1}}) \times (p_1 \cdot p_2) (p_3 \cdot p_4)
\end{align*}
With this term, we have the matrix product 
\[\varrho^{\gamma(3)}_{\delta(3)} \varrho^{\alpha(4)}_{\sigma(4)} (1- \varrho)^{\delta(1)}_{\beta(1)} (1- \vrho)^{\sigma(2)}_{\gamma(2)} = \Big[ \vrho_4 \cdot (1-\vrho_2) \cdot \vrho_3 \cdot  (1-\vrho_1) \Big]^\alpha_\beta \, . \]
\end{itemize}
We chose the compact notation $\vrho_k \equiv \vrho(p_k)$ for brevity, and used $\vrho_1 = \vrho_{\underline 1}$ thanks to the momentum-conserving function $\delta^{(3)}(\vp_1 - \vp_{\underline{1}})$.

Considering all terms in \eqref{eq:C11}, the scattering amplitude is always identical, and the matrix products arrange such that the final result has the expected ‘‘gain $-$ loss $+$ h.c.'' structure. Note that we considered here a particular ordering of the indices, while the full expression is symmetric through the exchange $(3,4,3',4')\leftrightarrow(4,3,4',3')$. In other words, one must take twice the previous result to account for all non-zero combinations.\footnote{This symmetry vanishes if $3$ and $4$ have the same flavour. However, this is precisely compensated by the extra factor of $2$ in the matrix elements for identical flavour, cf. table~\ref{Table:MatrixElements}.} Therefore,
\begin{equation}
\label{eq:C_nnscatt}
\begin{aligned}
\mathcal{C}^{[\nu\nu \leftrightarrow \nu \nu]} = &(2 \pi)^3 \delta^{(3)}(\vec{p}_1-\vec{p}_{\underline{1}}) \frac{2^5 G_F^2}{2}\int{[\dd^3 \vec{p}_2] [\dd^3 \vec{p}_3] [\dd^3 \vec{p}_4] (2 \pi)^4 \delta^{(4)}(p_1 + p_2 - p_3 - p_4)} \\
&\times (p_1 \cdot p_2)(p_3 \cdot p_4) \times F_\mathrm{sc}(\nu^{(1)},\nu^{(2)},\nu^{(3)},\nu^{(4)})
\end{aligned}
\end{equation}
with the statistical factor:
\begin{multline}
F_\mathrm{sc}(\nu^{(1)},\nu^{(2)},\nu^{(3)},\nu^{(4)}) =  \left[ \varrho_4 (1- \varrho_2) + \Tr(\cdots) \right] \varrho_3 (1-\varrho_1) + (1- \varrho_1) \varrho_3 \left[ (1- \varrho_2) \varrho_4 + \Tr(\cdots)\right]  \\
- \left[ (1- \varrho_4) \varrho_2  + \Tr(\cdots)\right] (1-\varrho_3)  \varrho_1 - \varrho_1  (1-\varrho_3)  \left[\varrho_2(1-\varrho_4)  + \Tr(\cdots)\right]  \, ,
\end{multline}
where $\Tr(\cdots)$ means the trace of the term in front of it.

Finally, the collision integral $\mathcal{I}$ which appears in the equation for $\vrho(p_1)$ is $\mathcal{C}$ without the momentum-conserving delta-function $\mathcal{C}[\vrho] = (2 \pi)^3 \, 2 E_1\, \delta^{(3)}(\vp_1 - \vp_{\underline{1}}) \mathcal{I}[\vrho]$.

\section{Energy conservation and QED equation of state}
\label{app:QED}

The transfer of entropy from electron/positron annihilations into the photon and neutrino baths is governed by the continuity equation $\dot{\rho} = - 3 H (\rho + P)$, which we rewrite as an equation on the dimensionless photon temperature $z(x)$ \cite{Mangano2002,Bennett2020}:
\begin{equation}
\frac{\dd z}{\dd x} =  \frac{\displaystyle \frac{x}{z}J(x/z) - \frac{1}{2 \pi^2 z^3} \frac{1}{xH} \int_{0}^{\infty}{\dd y \, y^3 \, \Tr \left[\mathcal{I}\right]} + G_1(x/z)}{ \displaystyle \frac{x^2}{z^2}J(x/z) + Y(x/z) + \frac{2 \pi^2}{15} + G_2(x/z)} \, , \label{eq:zQED}
\end{equation}
with
\begin{align}
J(\tau) &\equiv \frac{1}{\pi^2} \int_{0}^{\infty}{\dd \omega \, \omega^2 \frac{\exp{(\sqrt{\omega^2 + \tau^2})}}{(\exp{(\sqrt{\omega^2 + \tau^2})}+1)^2}} \, , \\
Y(\tau) &\equiv \frac{1}{\pi^2} \int_{0}^{\infty}{\dd \omega \, \omega^4 \frac{\exp{(\sqrt{\omega^2 + \tau^2})}}{(\exp{(\sqrt{\omega^2 + \tau^2})}+1)^2}} \, .
\end{align}
The $G_1$ and $G_2$ functions account for the modifications of the plasma equation of state due to finite-temperature QED corrections \cite{Heckler_PhRvD1994,Mangano2002,Bennett2020}. They can be calculated order by order in an expansion in powers of $\alpha = e^2/4\pi$. We use:
\begin{align}
G_1^{(2)}(\tau) &= 2 \pi \alpha \left[\frac{K'(\tau)}{3} + \frac{J'(\tau)}{6} + J'(\tau) K(\tau) + J(\tau) K'(\tau) \right] \, , \label{eq:G1e2} \\
G_2^{(2)}(\tau) &= - 8 \pi \alpha \left[\frac{K(\tau)}{6} + \frac{J(\tau)}{6} - \frac12 K(\tau)^2 + K(\tau)J(\tau)\right]  \nonumber \\
&\phantom{=}  + 2 \pi \alpha \tau \left[\frac{K'(\tau)}{6} - K(\tau)K'(\tau) + \frac{J'(\tau)}{6} + J'(\tau) K(\tau) + J(\tau) K'(\tau) \right] \, , \label{eq:G2e2} \\
G_1^{(3)}(\tau) &= - \sqrt{2 \pi} \alpha^{3/2} \sqrt{J(\tau)} \times \tau \left[2 j(\tau) - \tau j'(\tau) + \frac{\tau^2 j(\tau)^2}{2 J(\tau)} \right] \, , \label{eq:G1e3} \\
G_2^{(3)}(\tau) &= \sqrt{2 \pi} \alpha^{3/2} \sqrt{J(\tau)} \left[\frac{\left(2 J(\tau) + \tau^2 j(\tau)\right)^2}{2 J(\tau)} + 6 J(\tau) + \tau^2 \left( 3 j(\tau) - \tau j'(\tau)\right) \right] \, , \label{eq:G2e3}
\end{align}
where $(\cdots)' = \dd(\cdots)/\dd \tau$, and with the additional functions
\begin{align}
j(\tau) &\equiv \frac{1}{\pi^2} \int_{0}^{\infty}{\dd \omega \,  \frac{\exp{(\sqrt{\omega^2 + \tau^2})}}{(\exp{(\sqrt{\omega^2 + \tau^2})}+1)^2}} \, , \\
K(\tau) &\equiv \frac{1}{\pi^2} \int_{0}^{\infty}{\dd \omega \, \frac{\omega^2}{\sqrt{\omega^2 + \tau^2}} \frac{1}{\exp{(\sqrt{\omega^2 + \tau^2})}+1}} \, , \\
k(\tau) &\equiv \frac{1}{\pi^2} \int_{0}^{\infty}{\dd \omega \, \frac{1}{\sqrt{\omega^2 + \tau^2}} \frac{1}{\exp{(\sqrt{\omega^2 + \tau^2})}+1}} \, . 
\end{align}
We discarded a logarithmic contribution to $G_{1,2}^{(2)}$ that is subdominant compared to $G_{1,2}^{(3)}$ \cite{Bennett2020}. Note that our expressions are formally different from those of previous literature. For instance \eqref{eq:G1e2} is formally different from the one in \cite{Mangano2002,Bennett2020}, while \eqref{eq:G2e2} matches formally with \cite{Mangano2002}, but not with \cite{Bennett2020}. Finally, \eqref{eq:G1e3} and \eqref{eq:G2e3} slightly differ from expressions reported in \cite{Bennett2020}. Actually, all expressions are identical, since one can prove (after integrations by parts and rearrangements) the following identities:
\begin{equation}
J'(\tau) = - \tau j(\tau) \; , \
K'(\tau) = - \tau k(\tau) \; , \
Y'(\tau) = - 3 \tau J(\tau) \; , \
2 K(\tau) + \tau^2 k(\tau) = J(\tau) \, .
\end{equation}

\section{Quantum kinetic equations with antiparticles}
\label{app:antiparticles}

We present in this appendix the inclusion of antiparticles to the BBGKY formalism.

\paragraph{Generalized definitions.}

One must adapt the definitions \eqref{eq:defrhos2} and \eqref{eq:defHint} to include the annihilation and creation operators $\hb,\hbd$. Throughout this appendix, we will emphasize the indices which are associated to antiparticles with a barred notation $(\bar{\imath},\bar{\jmath})$. Therefore, with capital indices $I$ being either $i$ or $\bi$, we have:
\begin{align}
\label{eq:defrho_anti}
\varrho^{I_1 \cdots I_s}_{J_1 \cdots J_s} &\equiv \langle \hat{c}_{J_s}^\dagger \cdots \hat{c}_{J_1}^\dagger \hat{c}_{I_1} \cdots \hat{c}_{I_s} \rangle \, , \\
\hat{H}_0 &= \sum_{I,J}{t^{I}_{J} \, \hcd_I \hc_J} \, , \\
\hat{H}_{\rm int} &= \frac14 \sum_{I,J,K,L}{\tilde{v}^{IK}_{JL} \, \hcd_I \hcd_K \hc_L \hc_J} \, , \label{eq:vint_anti}
\end{align}
where $\hc_I = \ha_i$ or $\hb_{\bar{\imath}}$ depending on the index $I$ labelling a particle or an antiparticle.

The evolution equations \eqref{eq:hierarchy} and \eqref{eq:eqvrho} are naturally extended to the antiparticle case thanks to the global indices. The downside of this strategy is that the transformation law of tensors is now implicit: since $\ha$ transforms like $\hbd$ under a unitary transformation $\psi^a = \mathcal{U}^a_i \psi^i$, the behaviour of upper and lower indices is inverted whenever they label an antiparticle degree of freedom, for instance:
\begin{equation}
\label{eq:transfo_unit}
t^{i}_{j} = {\mathcal{U}^\dagger}^{i}_{a} \, t^{a}_{b} \, \mathcal{U}^{b}_{j} \qquad ; \qquad t^{\bi}_{\bj} = \mathcal{U}^{a}_{i} \, t^{\bar{a}}_{\bar{b}} \, {\mathcal{U}^\dagger}^{j}_b \, .
\end{equation}

Since we assume an isotropic medium, there are no ``abnormal'' or ``pairing'' densities \cite{Volpe_2013,SerreauVolpe,Volpe_2015} such as $\langle \hb \ha \rangle$, which ensures the separation of the two-body density matrix between the neutrino density matrix (for which we keep the notation $\vrho$) and the antineutrino one $\bvrho$. In order for $\bvrho$ to have the same transformation properties as $\vrho$, we need to take a transposed convention for its components:
\begin{equation}
\bvrho^\bi_\bj = \vrho^{\{J=\bar{\jmath}\}}_{\{I=\bar{\imath}\}} = \langle \hcd_\bi \hc_\bj \rangle = \langle \hbd_i \hb_j \rangle \, .
\end{equation}
One could further take transposed conventions for the antiparticle indices in $t$ and $\tilde{v}$, which would ensure a clear correspondence between index position and transformation law --- contrary to \eqref{eq:transfo_unit}. For instance, $\bar{t}^\bi_\bj \equiv t^\bj_\bi$ transforms as $t^i_j$. However, in order to keep a unique expression for the mean-field potential or the collision term, we stick to the general definitions above. For instance, we have:
\begin{equation}
\label{eq:Gamma_full}
\Gamma^{i}_{j} = \sum_{K,L}{\tilde{v}^{iK}_{jL} \vrho^{L}_{K}} = \sum_{k,l}{\tilde{v}^{ik}_{jl} \vrho^{l}_{k}} + \sum_{\bar{k},\bar{l}}{\tilde{v}^{i \bar{k}}_{j \bar{l}} \bvrho^{\bar{k}}_{\bar{l}}} \, .
\end{equation}
Since the annihilation and creation operators do not appear naturally in normal order in the Hamiltonian \eqref{eq:defHsm}, recasting it in the form  \eqref{eq:vint_anti} leads to extra minus signs in $\tilde{v}$ involving antiparticles (cf.~table~\ref{Table:MatrixElements}).

These conventions being settled, we can include the full set of interaction matrix elements and compute all relevant contributions to the neutrino QKEs \eqref{eq:QKE_rho}. In the following, we derive the QKE for $\bvrho$, which is not solved in this paper since we consider a zero asymmetry.\footnote{We just used the QKE for $\bvrho$ to check the numerical stability of the code.}

\paragraph{QKE for antineutrinos.}

Thanks to our conventions, the evolution equation for the antineutrino density matrix $\bvrho$ is similarly obtained within the BBGKY formalism, 
with some differences compared to the neutrino case. First and foremost, the evolution equation for $\bvrho^\bi_\bj$ correspond in the general formalism to the equation for $\vrho^\bj_\bi$:
	\begin{equation}
	\label{eq:drhobjbi}
	i  \frac{\dd \bvrho^{\bi}_{\bj}}{\dd t}  = i  \frac{\dd \vrho^{\bj}_{\bi}}{\dd t} = \left( \left[t^{\bj}_{K} + \Gamma^{\bj}_{K}\right] \vrho^{K}_{\bi} - \vrho^{\bj}_{K} \left[t^{K}_{\bi} + \Gamma^{K}_{\bi}\right] \right) + i \, \hat{\mathcal{C}}^\bj_\bi \, ,
	\end{equation}
showing that taking the commutator with a transposed convention leads to a minus sign. Moreover,
\begin{itemize}
	\item we express the kinetic terms $t^\bj_\bi$, starting from the mass basis:
	\begin{equation}
	 t^\bj_\bi = U^a_j  \left. \frac{\mathbb{M}^2}{2p}\right|^{\bar{a}}_{\bar{b}} {U^\dagger}^i_b = {U^\dagger}^i_b  \left. \frac{\mathbb{M}^2}{2p}\right|^b_a U^a_j = t^i_j \, ;
	 \end{equation}
	\item $\tilde{v}^{\bar{\jmath} k}_{\bar{\imath} l}$ is the coefficient in front of $\hbd_j \had_k \ha_l \hb_i$, so it will have the same expression (apart from the interchange of $u$ and $v$ spinors for neutrinos, which leaves the result identical) as the coefficient in front of $\ha_j \had_k \ha_l \had_i = - \had_i \had_k \ha_l \ha_j$, that is $- \tilde{v}^{ik}_{jl}$. Therefore, $\Gamma^{\bar{\jmath}}_{\bar{\imath}} = - \Gamma^{i}_{j}$.
\end{itemize}
Including these two results in \eqref{eq:drhobjbi} show that, compared to the neutrino case, the vacuum term gets a minus sign (from the reversed commutator), but not the mean-field. Formally,
\begin{equation}
i \frac{\dd \bvrho^i_j}{\dd t} = \left[- \hat{t} + \hat{\Gamma}, \hat{\bvrho}\right]^{i}_{j} + i \, \hat{\mathcal{C}}^\bj_\bi \, .
\end{equation} 
Two additional remarks:
\begin{itemize}
	\item $s$ and $t$ channels are inverted when the particle $1$ is an antineutrino ($2$ and $4$ left unchanged). For instance, the scattering between $\bnu_e$ and $e^-$ is a $s-$channel (exchanged momentum $\Delta = p_1+p_2$), contrary to the scattering between $\nu_e$ and $e^-$ ($\Delta = p_1 - p_2$). This changes the sign of $\Delta^2$, leading to another minus sign for $\Gamma$ at order $1/m_{W,Z}^2$;
	\item the collision integral $\bar{\mathcal{I}}$ is obtained from $\mathcal{I}$ through the replacements $\vrho \leftrightarrow \bvrho$ and $g_L \leftrightarrow g_R$.
\end{itemize}
Considering all these remarks, we obtained the QKE for $\bvrho$ \eqref{eq:QKE_rhobar}.

\section{Effect of the CP violating phase}
\label{app:CP}

The generalized parametrization of the PMNS matrix \eqref{eq:PMNS} when including a CP violating phase\footnote{We do not include possible Majorana phases that have no effect on neutrino oscillations.} reads
\begin{equation}
\label{eq:PMNS_CP}
U = R_{23} S R_{13} S^\dagger R_{12} = \begin{pmatrix} 
c_{12} c_{13} & s_{12} c_{13} & s_{13} e^{-i \delta} \\
- s_{12}c_{23} - c_{12}s_{23}s_{13} e^{i \delta} & c_{12} c_{23} - s_{12}s_{23}s_{13} e^{i \delta} & s_{23} c_{13} \\
s_{12}s_{23} - c_{12}c_{23}s_{13} e^{i \delta} & -c_{12}s_{23} - s_{12}c_{23}s_{13} e^{i \delta} & c_{23} c_{13}
\end{pmatrix} \, ,
\end{equation}
where $S = \mathrm{diag}(1,1,e^{i \delta})$. 

Although this new phase affects the vacuum oscillation term in the QKEs, it is actually possible to factorise this dependence and reduce the problem to the case $\delta = 0$, in some limits that we expose below. We follow the derivation of refs. \cite{Balantekin:2007es,Gava:2008rp,Gava:2010kz,Gava_corr}, where conditions under which the CP phase has an impact on the evolution of $\vrho$ in matter were first uncovered.

In this section, we will note with a superscript $^0$ the quantities in the $\delta =0$ case. We introduce a convenient unitary transformation $\check{S} \equiv R_{23} S R_{23}^\dagger$ and define $\check{\vrho} \equiv \check{S}^\dagger \vrho \check{S}$ (likewise for $\bvrho$). Let us now prove that $\check{\vrho} = \vrho^0$ \cite{Gava:2010kz}. First, we need to show that $\check{\vrho}$ has the same evolution equation as $\vrho^0$. Let us rewrite the QKE \eqref{eq:QKE_final} in a very compact way:
\begin{equation}
\label{eq:QKE_compact_CP}
i \frac{\partial \vrho}{\partial x} =  \lambda [U \mathbb{M}^2 U^\dagger,\vrho] + \mu [\bar{\mathbb{E}}_e + \bar{\mathbb{P}}_e, \vrho]  + i \mathcal{K}[\vrho,\bvrho] \, ,
\end{equation}
with coefficients $\lambda, \mu$ which can be read from \eqref{eq:QKE_final}. Applying $\check{S}^\dagger (\cdots) \check{S}$ on both sides of the QKE gives the evolution equation for $\check{\vrho}$. 

First, using that $\check{S}^\dagger U = U^0 S^\dagger$ (we recall that $U^0$ is the PMNS matrix without CP phase) and that $\mathbb{M}^2$ and $S$ commute since they are diagonal, the vacuum term reads $\check{S}^\dagger [U \mathbb{M}^2 U^\dagger,\vrho] \check{S} = [U^0 \mathbb{M}^2 {U^0}^\dagger, \check{\vrho}] $. Then, the mean-field term satisfies $\check{S}^\dagger [\bar{\mathbb{E}}_e + \bar{\mathbb{P}}_e, \vrho] \check{S} = [\bar{\mathbb{E}}_e + \bar{\mathbb{P}}_e, \check{\vrho}]$. This property only holds because the energy density of muons is negligible, ensuring that muon and tau neutrinos have the same interactions \cite{Gava:2010kz}. Finally, the collision term contains products of density matrices and $G_{L,R}$ coupling matrices for the scattering/annihilation terms with electrons and positrons. Since $[G_{L,R}, \check{S}^{(\dagger)}]=0$, we can write $\check{S}^\dagger \mathcal{K}[\vrho, \bvrho] \check{S} = \mathcal{K}[\check{\vrho},\check{\bvrho}]$. Once again, the fact that $\nu_\mu$ and $\nu_\tau$ have identical interactions is key to this factorisation, as pointed out in ref.~\cite{Gava:2010kz} and previously in refs.~\cite{Balantekin:2007es,Gava:2008rp} in the astrophysical context. In ref.~\cite{Gava:2010kz}, the collision term is approximated by a damping factor; the factorisation then holds since the damping coefficients are identical whether they involve $\nu_\mu$ or $\nu_\tau$.

All in all, the QKE for $\check{\vrho}$ reads:
\begin{equation}
\label{eq:QKE_compact_CP_S}
i \frac{\partial \check{\vrho}}{\partial x} =  \lambda [U^0 \mathbb{M}^2 {U^0}^\dagger,\check{\vrho}] + \mu [\bar{\mathbb{E}}_e + \bar{\mathbb{P}}_e, \check{\vrho}]  + i \mathcal{K}[\check{\vrho},\check{\bvrho}] \, ,
\end{equation}
which is exactly the QKE for $\vrho^0$, i.e., the QKE without CP phase. Moreover, the initial condition \eqref{eq:initial_condition} is unaffected by the $\check{S}$ transformation: $\check{\vrho}(x_{\rm in},y) = \vrho^0(x_{\rm in},y)$. Since the initial conditions and the evolution equations are identical for $\check{\vrho}$ and $\vrho^0$, then at all times $\vrho^0(x,y) = \check{\vrho}(x,y)$ \cite{Gava:2010kz,Balantekin:2007es}. We can therefore write the relation between the density matrices with and without CP phase,
\begin{equation}
\label{eq:CP_flavour}
\vrho(x,y) = \check{S} \vrho^0(x,y) \check{S}^\dagger \, .
\end{equation}
This relation has two major consequences:
\begin{enumerate}
	\item The trace of $\vrho$ is unaffected by $\delta$, therefore $\Neff = \Neff(\delta = 0)$;
	\item The first diagonal component is unchanged $\vrho^e_e = (\vrho^0)^e_e$. Equivalently with the parametrization \eqref{eq:param_rho}, $z_{\nu_e} = z_{\nu_e}^0$ and $\delta g_{\nu_e} = \delta g_{\nu_e}^0$.
\end{enumerate}
Therefore, under the assumptions made above (in particular, the initial distribution has no chemical potentials), the CP phase will have no effect on BBN, since light element abundances are only sensitive to $\Neff$, $z_{\nu_e}$ and $\delta g_{\nu_e}$. Note that in presence of initial degeneracies, the initial conditions do not necessarily coincide $\check{\vrho}(x_{\rm in},y) \neq \vrho^0(x_{\rm in},y)$ and signatures of a CP phase could be found in the primordial abundances \cite{Gava:2010kz,Gava_corr}. 

A useful rewriting of \eqref{eq:CP_flavour} can be made with the final distributions ($x = x_f$), when mean-field effects are negligible. The correspondence between $\delta =0$ and $\delta \neq 0$ then reads in the matter basis (which is then the mass basis):
\begin{align}
\vrho_m(x_f,y) &= S \vrho_m^0(x_f,y) S^\dagger \ .  
\\
  \intertext{Note that the transformation involves now $S$ instead of $\check{S}$ (this is linked to the fact that the transformation between $\vrho$ and $\vrho_m$ is made through $U$, while the transformation between $\vrho^0$ and $\vrho_m^0$ involves $U^0$). We can go further using the ATAO approximation, which constrains the form of $\vrho$ and allows to analytically estimate the effect of the CP phase. \emph{In the ATAO approximation}, $\vrho_m$ is diagonal, such that we get the result:} 
  \vrho_m(x_f,y) &= \vrho_m^0(x_f,y) \qquad \text{\textbf{[ATAO]}}
\end{align}
Defining effective temperatures $z_{\nu_i}$ for the mass states ($i=1,2,3$), we have then $z_{\nu_i} = z_{\nu_i}^0$. Using the PMNS matrix to express the results in the flavour basis, the effective temperatures read:\footnote{These expressions are rigorously exact for the energy densities, and they can be rewritten for the effective temperatures since $z_\nu - 1 \ll 1$.}
\begin{equation}
\label{eq:resultsCP}
\begin{aligned}
z_{\nu_e} &= z_{\nu_e}^0 \ , \\
z_{\nu_\mu} &= z_{\nu_\mu}^0 - \frac12 (z_{\nu_1} - z_{\nu_2}) \sin{(2 \theta_{12})} \sin{(\theta_{13})}\sin{(2 \theta_{23})} \left[ 1 - \cos{(\delta)}\right] \ , \\
z_{\nu_\tau} &= z_{\nu_\tau}^0 + \frac12 (z_{\nu_1} - z_{\nu_2}) \sin{(2 \theta_{12})} \sin{(\theta_{13})}\sin{(2 \theta_{23})} \left[ 1 - \cos{(\delta)}\right] \ .
\end{aligned}
\end{equation}
These relations show that the CP phase only affects the muon and tau neutrino distribution functions, with a $[\cos{(\delta)} - 1]$ dependence. For the preferred values of $\delta = 1.36 \, \pi$ and the mixing angles \cite{PDG}, and with the results for $\delta=0$ from Section~\ref{subsec:results}, we expect $\abs{z_{\nu_\mu} - z_{\nu_\mu}^0} =  \abs{z_{\nu_\tau} - z_{\nu_\tau}^0} \simeq 4.6 \times 10^{-5}$.  This is in excellent agreement with the numerical results obtained solving the QKE with a CP phase (table~\ref{Table:Res_NuDec_CP}).

\renewcommand{\arraystretch}{1.2}

\begin{table}[!htb]
	\centering
	\begin{tabular}{|l|ccccc|}
  	\hline 
  Final values & $z$ & $z_{\nu_e}$  & $z_{\nu_\mu}$ & $z_{\nu_\tau}$ &$\Neff$  \\
  \hline \hline
    $\delta = 0$ & $1.39797$ & $1.00175$ & $1.00132$ & $1.00130$ & $3.04397$   \\
    $\delta = 1.36 \, \pi$  & $1.39797$ & $1.00175$ & $1.00127$ & $1.00135$ & $3.04397$   \\ \hline
\end{tabular}
	\caption{Frozen-out values of the dimensionless photon and neutrino temperatures, and the effective number of neutrino species. We compare the results without CP phase (see also table~\ref{Table:Res_NuDec}) and with the average value for $\delta$ from \cite{PDG}.
	\label{Table:Res_NuDec_CP}}
\end{table}

Finally, the antineutrino density matrices satisfy the same relation as for neutrinos \eqref{eq:CP_flavour} $\bvrho(x,y) = \check{S} \bvrho^0(x,y) \check{S}^\dagger$. The QKEs in the absence of CP phase preserve the property $\vrho^0 = {\bvrho^0}^*$ if it is true initially. The asymmetry with $\delta \neq 0$ would then read $\vrho - \bvrho = \check{S}(\vrho^0 - {\vrho^0}^*)\check{S}^\dagger$. Therefore, CP violation effects in the $\nu_\mu$ and $\nu_\tau$ distributions (which would be contributions $\propto \sin{\delta}$) can arise from the complex components of $\vrho^0$, thus requiring the ATAO approximation to break down. Since in the cosmological context without initial degeneracies the approximation is very well satisfied, there can be no additional CP violation and the formulae \eqref{eq:resultsCP} are equally valid for antineutrinos.

\bibliographystyle{JHEP}
\bibliography{BiblioAdiabBBGKY}

\providecommand{\href}[2]{#2}\begingroup\raggedright\begin{thebibliography}{10}

\bibitem{Dodelson_Turner_PhRvD1992}
S.~{Dodelson} and M.~S. {Turner}, \emph{{Nonequilibrium neutrino statistical
  mechanics in the expanding Universe}},
  \href{https://doi.org/10.1103/PhysRevD.46.3372}{\emph{Phys. Rev. D}
  {\bfseries 46} (1992) 3372}.

\bibitem{Dolgov1992}
A.~Dolgov and M.~Fukugita, \emph{{Nonequilibrium effect of the neutrino
  distribution on primordial helium synthesis}},
  \href{https://doi.org/10.1103/PhysRevD.46.5378}{\emph{Phys. Rev. D}
  {\bfseries 46} (1992) 5378}.

\bibitem{Dolgov_NuPhB1997}
A.~D. Dolgov, S.~H. Hansen and D.~V. Semikoz, \emph{{Nonequilibrium corrections
  to the spectra of massless neutrinos in the early universe}},
  \href{https://doi.org/10.1016/S0550-3213(97)00479-3}{\emph{Nucl. Phys. B}
  {\bfseries 503} (1997) 426}
  [\href{https://arxiv.org/abs/hep-ph/9703315}{{\ttfamily hep-ph/9703315}}].

\bibitem{Esposito_NuPhB2000}
S.~{Esposito}, G.~{Miele}, S.~{Pastor}, M.~{Peloso} and O.~{Pisanti},
  \emph{{Non equilibrium spectra of degenerate relic neutrinos}},
  \href{https://doi.org/10.1016/S0550-3213(00)00554-X}{\emph{Nucl. Phys. B}
  {\bfseries 590} (2000) 539}
  [\href{https://arxiv.org/abs/astro-ph/0005573}{{\ttfamily
  astro-ph/0005573}}].

\bibitem{Mangano2002}
G.~{Mangano}, G.~{Miele}, S.~{Pastor} and M.~{Peloso}, \emph{{A precision
  calculation of the effective number of cosmological neutrinos}},
  \href{https://doi.org/10.1016/S0370-2693(02)01622-2}{\emph{Phys. Lett. B}
  {\bfseries 534} (2002) 8}
  [\href{https://arxiv.org/abs/astro-ph/0111408}{{\ttfamily
  astro-ph/0111408}}].

\bibitem{Grohs2015}
E.~Grohs, G.~M. Fuller, C.~T. Kishimoto, M.~W. Paris and A.~Vlasenko,
  \emph{Neutrino energy transport in weak decoupling and big bang
  nucleosynthesis},
  \href{https://doi.org/10.1103/PhysRevD.93.083522}{\emph{Phys. Rev. D}
  {\bfseries 93} (2016) 083522}
  [\href{https://arxiv.org/abs/1512.02205}{{\ttfamily 1512.02205}}].

\bibitem{Froustey2019}
J.~Froustey and C.~Pitrou, \emph{Incomplete neutrino decoupling effect on big
  bang nucleosynthesis},
  \href{https://doi.org/10.1103/PhysRevD.101.043524}{\emph{Phys. Rev. D}
  {\bfseries 101} (2020) 043524}
  [\href{https://arxiv.org/abs/1912.09378}{{\ttfamily 1912.09378}}].

\bibitem{Heckler_PhRvD1994}
A.~F. {Heckler}, \emph{{Astrophysical applications of quantum corrections to
  the equation of state of a plasma}},
  \href{https://doi.org/10.1103/PhysRevD.49.611}{\emph{Phys. Rev. D} {\bfseries
  49} (1994) 611}.

\bibitem{Bennett2020}
J.~J. Bennett, G.~Buldgen, M.~Drewes and Y.~Y. Wong, \emph{{Towards a precision
  calculation of the effective number of neutrinos $N_{\rm eff}$ in the
  Standard Model I: The QED equation of state}},
  \href{https://doi.org/10.1088/1475-7516/2020/03/003}{\emph{JCAP} {\bfseries
  03} (2020) 003} [\href{https://arxiv.org/abs/1911.04504}{{\ttfamily
  1911.04504}}].

\bibitem{GiuntiKim}
C.~Giunti and C.~W. Kim, \emph{{Fundamentals of Neutrino Physics and
  Astrophysics}}. Oxford University Press, 2007.

\bibitem{Mangano2005}
G.~{Mangano}, G.~{Miele}, S.~{Pastor}, T.~{Pinto}, O.~{Pisanti} and P.~D.
  {Serpico}, \emph{{Relic neutrino decoupling including flavour oscillations}},
  \href{https://doi.org/10.1016/j.nuclphysb.2005.09.041}{\emph{Nucl. Phys. B}
  {\bfseries 729} (2005) 221}
  [\href{https://arxiv.org/abs/hep-ph/0506164}{{\ttfamily hep-ph/0506164}}].

\bibitem{Relic2016_revisited}
P.~F. de~Salas and S.~Pastor, \emph{{Relic neutrino decoupling with flavour
  oscillations revisited}},
  \href{https://doi.org/10.1088/1475-7516/2016/07/051}{\emph{JCAP} {\bfseries
  07} (2016) 051} [\href{https://arxiv.org/abs/1606.06986}{{\ttfamily
  1606.06986}}].

\bibitem{Gariazzo_2019}
S.~Gariazzo, P.~de~Salas and S.~Pastor, \emph{{Thermalisation of sterile
  neutrinos in the early Universe in the 3+1 scheme with full mixing matrix}},
  \href{https://doi.org/10.1088/1475-7516/2019/07/014}{\emph{JCAP} {\bfseries
  07} (2019) 014} [\href{https://arxiv.org/abs/1905.11290}{{\ttfamily
  1905.11290}}].

\bibitem{Akita2020}
K.~Akita and M.~Yamaguchi, \emph{{A precision calculation of relic neutrino
  decoupling}},
  \href{https://doi.org/10.1088/1475-7516/2020/08/012}{\emph{JCAP} {\bfseries
  08} (2020) 012} [\href{https://arxiv.org/abs/2005.07047}{{\ttfamily
  2005.07047}}].

\bibitem{Escudero_2018}
M.~Escudero, \emph{{Neutrino decoupling beyond the Standard Model: CMB
  constraints on the Dark Matter mass with a fast and precise $N_{\rm eff}$
  evaluation}},
  \href{https://doi.org/10.1088/1475-7516/2019/02/007}{\emph{JCAP} {\bfseries
  02} (2019) 007} [\href{https://arxiv.org/abs/1812.05605}{{\ttfamily
  1812.05605}}].

\bibitem{Escudero_2020}
M.~Escudero, \emph{{Precision early universe thermodynamics made simple:
  $N_{\rm eff}$ and neutrino decoupling in the Standard Model and beyond}},
  \href{https://doi.org/10.1088/1475-7516/2020/05/048}{\emph{JCAP} {\bfseries
  05} (2020) 048} [\href{https://arxiv.org/abs/2001.04466}{{\ttfamily
  2001.04466}}].

\bibitem{Pitrou_2018PhysRept}
C.~Pitrou, A.~Coc, J.-P. Uzan and E.~Vangioni, \emph{Precision big bang
  nucleosynthesis with improved helium-4 predictions},
  \href{https://doi.org/https://doi.org/10.1016/j.physrep.2018.04.005}{\emph{Phys.
  Rept.} {\bfseries 754} (2018) 1}
  [\href{https://arxiv.org/abs/1801.08023}{{\ttfamily 1801.08023}}].

\bibitem{Planck18}
{\scshape Planck} collaboration, \emph{{Planck 2018 results. VI. Cosmological
  parameters}},
  \href{https://doi.org/10.1051/0004-6361/201833910}{\emph{Astron. Astrophys.}
  {\bfseries 641} (2020) A6}
  [\href{https://arxiv.org/abs/1807.06209}{{\ttfamily 1807.06209}}].

\bibitem{Kreisch:2019yzn}
C.~D. Kreisch, F.-Y. Cyr-Racine and O.~Dor{\'e}, \emph{{The Neutrino Puzzle:
  Anomalies, Interactions, and Cosmological Tensions}},
  \href{https://doi.org/10.1103/PhysRevD.101.123505}{\emph{Phys. Rev. D}
  {\bfseries 101} (2020) 123505}
  [\href{https://arxiv.org/abs/1902.00534}{{\ttfamily 1902.00534}}].

\bibitem{Grohs:2020xxd}
E.~Grohs, G.~M. Fuller and M.~Sen, \emph{{Consequences of neutrino self
  interactions for weak decoupling and big bang nucleosynthesis}},
  \href{https://doi.org/10.1088/1475-7516/2020/07/001}{\emph{JCAP} {\bfseries
  07} (2020) 001} [\href{https://arxiv.org/abs/2002.08557}{{\ttfamily
  2002.08557}}].

\bibitem{SiglRaffelt}
G.~{Sigl} and G.~{Raffelt}, \emph{{General kinetic description of relativistic
  mixed neutrinos}},
  \href{https://doi.org/10.1016/0550-3213(93)90175-O}{\emph{Nucl. Phys. B}
  {\bfseries 406} (1993) 423}.

\bibitem{Vlasenko:2013fja}
A.~Vlasenko, G.~M. Fuller and V.~Cirigliano, \emph{{Neutrino Quantum
  Kinetics}}, \href{https://doi.org/10.1103/PhysRevD.89.105004}{\emph{Phys.
  Rev. D} {\bfseries 89} (2014) 105004}
  [\href{https://arxiv.org/abs/1309.2628}{{\ttfamily 1309.2628}}].

\bibitem{BlaschkeCirigliano}
D.~N. Blaschke and V.~Cirigliano, \emph{Neutrino quantum kinetic equations: The
  collision term},
  \href{https://doi.org/10.1103/PhysRevD.94.033009}{\emph{Phys. Rev. D}
  {\bfseries 94} (2016) 033009}
  [\href{https://arxiv.org/abs/1605.09383}{{\ttfamily 1605.09383}}].

\bibitem{Volpe_2013}
C.~{Volpe}, D.~{V{\"a}{\"a}n{\"a}nen} and C.~{Espinoza}, \emph{{Extended
  evolution equations for neutrino propagation in astrophysical and
  cosmological environments}},
  \href{https://doi.org/10.1103/PhysRevD.87.113010}{\emph{Phys. Rev. D}
  {\bfseries 87} (2013) 113010}
  [\href{https://arxiv.org/abs/1302.2374}{{\ttfamily 1302.2374}}].

\bibitem{Volpe_2015}
C.~{Volpe}, \emph{{Neutrino quantum kinetic equations}},
  \href{https://doi.org/10.1142/S0218301315410098}{\emph{Int. J. Mod. Phys. E}
  {\bfseries 24} (2015) 1541009}
  [\href{https://arxiv.org/abs/1506.06222}{{\ttfamily 1506.06222}}].

\bibitem{SerreauVolpe}
J.~Serreau and C.~Volpe, \emph{Neutrino-antineutrino correlations in dense
  anisotropic media},
  \href{https://doi.org/10.1103/PhysRevD.90.125040}{\emph{Phys. Rev. D}
  {\bfseries 90} (2014) 125040}
  [\href{https://arxiv.org/abs/1409.3591}{{\ttfamily 1409.3591}}].

\bibitem{Gava:2010kz}
J.~Gava and C.~Volpe, \emph{{CP violation effects on the neutrino degeneracy
  parameters in the Early Universe}},
  \href{https://doi.org/10.1016/j.nuclphysb.2010.04.024}{\emph{Nucl. Phys. B}
  {\bfseries 837} (2010) 50} [\href{https://arxiv.org/abs/1002.0981}{{\ttfamily
  1002.0981}}].

\bibitem{Gava_corr}
M.~C. Volpe, \emph{{Corrigendum to “CP violation effects on the neutrino
  degeneracy parameters in the Early Universe” [Nucl. Phys. B 837 (2010)
  50–60)]}},
  \href{https://doi.org/https://doi.org/10.1016/j.nuclphysb.2020.115035}{\emph{Nucl.
  Phys. B} {\bfseries 957} (2020) 115035}.

\bibitem{Cassing1990}
W.~Cassing and U.~Mosel, \emph{{Many body theory of high-energy heavy ion
  reactions}}, \href{https://doi.org/10.1016/0146-6410(90)90032-Y}{\emph{Prog.
  Part. Nucl. Phys.} {\bfseries 25} (1990) 235}.

\bibitem{Reinhard1994}
P.-G. Reinhard and C.~Toepffer, \emph{Correlations in nuclei and nuclear
  dynamics},
  \href{https://doi.org/https://doi.org/10.1142/S0218301394000139}{\emph{Int.
  J. Mod. Phys. E} {\bfseries 3} (1994) 435}.

\bibitem{Lac04}
D.~Lacroix, S.~Ayik and P.~Chomaz, \emph{Nuclear collective vibrations in
  extended mean-field theory},
  \href{https://doi.org/https://doi.org/10.1016/j.ppnp.2004.02.002}{\emph{Prog.
  Part. Nucl. Phys.} {\bfseries 52} (2004) 497}.

\bibitem{Simenel}
C.~{Simenel}, B.~{Avez} and D.~{Lacroix}, \emph{{Microscopic approaches for
  nuclear Many-Body dynamics: applications to nuclear reactions}},
  \href{https://arxiv.org/abs/0806.2714}{{\ttfamily 0806.2714}}.

\bibitem{Lac14}
D.~{Lacroix} and S.~{Ayik}, \emph{{Stochastic quantum dynamics beyond mean
  field}}, \href{https://doi.org/10.1140/epja/i2014-14095-8}{\emph{Eur. Phys.
  J. A} {\bfseries 50} (2014) 95}
  [\href{https://arxiv.org/abs/1402.2393}{{\ttfamily 1402.2393}}].

\bibitem{Bogoliubov}
N.~Bogoliubov, \emph{{Kinetic equations}}, {\emph{Journal of Physics USSR}
  {\bfseries 10} (1946) 265}.

\bibitem{BornGreen}
M.~Born and H.~Green, \emph{{A General Kinetic Theory of Liquids. I. The
  Molecular Distribution Functions}},
  \href{https://doi.org/10.1098/rspa.1946.0093}{\emph{Proc. Roy. Soc. Lond. A}
  {\bfseries A188} (1946) 10}.

\bibitem{Kirkwood}
J.~G. {Kirkwood}, \emph{{The Statistical Mechanical Theory of Transport
  Processes I. General Theory}},
  \href{https://doi.org/10.1063/1.1724117}{\emph{J. Chem. Phys.} {\bfseries 14}
  (1946) 180}.

\bibitem{Yvon}
J.~Yvon, \emph{La th{\'e}orie statistique des fluides et l'{\'e}quation
  d'{\'e}tat}, vol.~203 of \emph{Actualit{\'e}s scientifiques et
  industrielles}. Hermann \& cie, 1935.

\bibitem{NotzoldRaffelt_NuPhB1988}
D.~{N{\"o}tzold} and G.~{Raffelt}, \emph{{Neutrino dispersion at finite
  temperature and density}},
  \href{https://doi.org/10.1016/0550-3213(88)90113-7}{\emph{Nucl. Phys. B}
  {\bfseries 307} (1988) 924}.

\bibitem{Grohs2019}
L.~C. Thomas, T.~Dezen, E.~B. Grohs and C.~T. Kishimoto,
  \emph{Electron-positron annihilation freeze-out in the early universe},
  \href{https://doi.org/10.1103/PhysRevD.101.063507}{\emph{Phys. Rev. D}
  {\bfseries 101} (2020) 063507}
  [\href{https://arxiv.org/abs/1910.14050}{{\ttfamily 1910.14050}}].

\bibitem{Hannestad_PhRvD1995}
S.~{Hannestad} and J.~{Madsen}, \emph{{Neutrino decoupling in the early
  Universe}}, \href{https://doi.org/10.1103/PhysRevD.52.1764}{\emph{Phys. Rev.
  D} {\bfseries 52} (1995) 1764}
  [\href{https://arxiv.org/abs/astro-ph/9506015}{{\ttfamily
  astro-ph/9506015}}].

\bibitem{Semikoz_Tkachev}
D.~Semikoz and I.~Tkachev, \emph{{Condensation of bosons in kinetic regime}},
  \href{https://doi.org/10.1103/PhysRevD.55.489}{\emph{Phys. Rev. D} {\bfseries
  55} (1997) 489} [\href{https://arxiv.org/abs/hep-ph/9507306}{{\ttfamily
  hep-ph/9507306}}].

\bibitem{HannestadTamborra}
S.~Hannestad, I.~Tamborra and T.~Tram, \emph{{Thermalisation of light sterile
  neutrinos in the early universe}},
  \href{https://doi.org/10.1088/1475-7516/2012/07/025}{\emph{JCAP} {\bfseries
  07} (2012) 025} [\href{https://arxiv.org/abs/1204.5861}{{\ttfamily
  1204.5861}}].

\bibitem{MSW_W}
L.~Wolfenstein, \emph{{Neutrino Oscillations in Matter}},
  \href{https://doi.org/10.1103/PhysRevD.17.2369}{\emph{Phys. Rev. D}
  {\bfseries 17} (1978) 2369}.

\bibitem{MSW_MS}
S.~Mikheyev and A.~Smirnov, \emph{{Resonance Amplification of Oscillations in
  Matter and Spectroscopy of Solar Neutrinos}}, {\emph{Sov. J. Nucl. Phys.}
  {\bfseries 42} (1985) 913}.

\bibitem{ODEPACK}
A.~C. Hindmarsh, \emph{{ODEPACK, A Systematized Collection of ODE Solvers}},
  {\emph{IMACS Transactions on Scientific Computation} {\bfseries 1} (1983)
  55}.

\bibitem{Dolgov_NuPhB1999}
A.~D. {Dolgov}, S.~H. {Hansen} and D.~V. {Semikoz}, \emph{{Non-equilibrium
  corrections to the spectra of massless neutrinos in the early universe}},
  \href{https://doi.org/10.1016/S0550-3213(98)00818-9}{\emph{Nucl. Phys. B}
  {\bfseries 543} (1999) 269}
  [\href{https://arxiv.org/abs/hep-ph/9805467}{{\ttfamily hep-ph/9805467}}].

\bibitem{PDG}
{\scshape Particle Data Group} collaboration, \emph{{Review of Particle
  Physics}}, \href{https://doi.org/10.1093/ptep/ptaa104}{\emph{Prog. Theor.
  Exp. Phys.} {\bfseries 2020} (2020) 083C01}.

\bibitem{Passera_QED}
M.~Passera, \emph{{QED corrections to neutrino-electron scattering}},
  \href{https://doi.org/10.1103/PhysRevD.64.113002}{\emph{Phys. Rev. D}
  {\bfseries 64} (2001) 113002}
  [\href{https://arxiv.org/abs/hep-ph/0011190}{{\ttfamily hep-ph/0011190}}].

\bibitem{Esposito_QED}
S.~Esposito, G.~Mangano, G.~Miele, I.~Picardi and O.~Pisanti, \emph{{Neutrino
  energy loss rate in a stellar plasma}},
  \href{https://doi.org/10.1016/S0550-3213(03)00151-2}{\emph{Nucl. Phys. B}
  {\bfseries 658} (2003) 217}
  [\href{https://arxiv.org/abs/astro-ph/0301438}{{\ttfamily
  astro-ph/0301438}}].

\bibitem{Damping98}
N.~F. Bell, R.~R. Volkas and Y.~Y.~Y. Wong, \emph{Relic neutrino asymmetry
  evolution from first principles},
  \href{https://doi.org/10.1103/PhysRevD.59.113001}{\emph{Phys. Rev. D}
  {\bfseries 59} (1999) 113001}
  [\href{https://arxiv.org/abs/hep-ph/9809363}{{\ttfamily hep-ph/9809363}}].

\bibitem{Dolgov_NuPhB2002}
A.~D. {Dolgov}, S.~H. {Hansen}, S.~{Pastor}, S.~T. {Petcov}, G.~G. {Raffelt}
  and D.~V. {Semikoz}, \emph{{Cosmological bounds on neutrino degeneracy
  improved by flavor oscillations}},
  \href{https://doi.org/10.1016/S0550-3213(02)00274-2}{\emph{Nucl. Phys. B}
  {\bfseries 632} (2002) 363}
  [\href{https://arxiv.org/abs/hep-ph/0201287}{{\ttfamily hep-ph/0201287}}].

\bibitem{Mirizzi2012}
A.~Mirizzi, N.~Saviano, G.~Miele and P.~D. Serpico, \emph{Light sterile
  neutrino production in the early universe with dynamical neutrino
  asymmetries}, \href{https://doi.org/10.1103/PhysRevD.86.053009}{\emph{Phys.
  Rev. D} {\bfseries 86} (2012) 053009}
  [\href{https://arxiv.org/abs/1206.1046}{{\ttfamily 1206.1046}}].

\bibitem{Saviano2013}
N.~Saviano, A.~Mirizzi, O.~Pisanti, P.~D. Serpico, G.~Mangano and G.~Miele,
  \emph{Multimomentum and multiflavor active-sterile neutrino oscillations in
  the early universe: Role of neutrino asymmetries and effects on
  nucleosynthesis},
  \href{https://doi.org/10.1103/PhysRevD.87.073006}{\emph{Phys. Rev. D}
  {\bfseries 87} (2013) 073006}
  [\href{https://arxiv.org/abs/1302.1200}{{\ttfamily 1302.1200}}].

\bibitem{CMB-S4}
K.~Abazajian et~al., \emph{{CMB-S4 Decadal Survey APC White Paper}},
  \href{https://doi.org/10.2172/1556957}{\emph{Bull. Am. Astron. Soc.}
  {\bfseries 51} (2019) 209}
  [\href{https://arxiv.org/abs/1908.01062}{{\ttfamily 1908.01062}}].

\bibitem{Fields_PhRvD1993}
B.~D. {Fields}, S.~{Dodelson} and M.~S. {Turner}, \emph{{Effect of neutrino
  heating on primordial nucleosynthesis}},
  \href{https://doi.org/10.1103/PhysRevD.47.4309}{\emph{Phys. Rev. D}
  {\bfseries 47} (1993) 4309}
  [\href{https://arxiv.org/abs/astro-ph/9210007}{{\ttfamily
  astro-ph/9210007}}].

\bibitem{SmithBBN}
M.~S. Smith, L.~H. Kawano and R.~A. Malaney, \emph{{Experimental,
  computational, and observational analysis of primordial nucleosynthesis}},
  \href{https://doi.org/10.1086/191763}{\emph{Astrophys. J. Suppl.} {\bfseries
  85} (1993) 219}.

\bibitem{PitrouPospelov20}
C.~{Pitrou} and M.~{Pospelov}, \emph{{QED corrections to Big-Bang
  nucleosynthesis reaction rates}},
  \href{https://doi.org/10.1103/PhysRevC.102.015803}{\emph{Phys. Rev. C}
  {\bfseries 102} (2020) 015803}
  [\href{https://arxiv.org/abs/1904.07795}{{\ttfamily 1904.07795}}].

\bibitem{Parthenope}
O.~{Pisanti}, A.~{Cirillo}, S.~{Esposito}, F.~{Iocco}, G.~{Mangano}, G.~{Miele}
  et~al., \emph{{PArthENoPE: Public algorithm evaluating the nucleosynthesis of
  primordial elements}},
  \href{https://doi.org/10.1016/j.cpc.2008.02.015}{\emph{Comput. Phys. Commun.}
  {\bfseries 178} (2008) 956}
  [\href{https://arxiv.org/abs/0705.0290}{{\ttfamily 0705.0290}}].

\bibitem{Parthenope_reloaded}
R.~{Consiglio}, P.~F. {de Salas}, G.~{Mangano}, G.~{Miele}, S.~{Pastor} and
  O.~{Pisanti}, \emph{{PArthENoPE reloaded}},
  \href{https://doi.org/10.1016/j.cpc.2018.06.022}{\emph{Comput. Phys. Commun.}
  {\bfseries 233} (2018) 237}
  [\href{https://arxiv.org/abs/1712.04378}{{\ttfamily 1712.04378}}].

\bibitem{FidlerPitrou}
C.~Fidler and C.~Pitrou, \emph{{Kinetic theory of fermions in curved
  spacetime}}, \href{https://doi.org/10.1088/1475-7516/2017/06/013}{\emph{JCAP}
  {\bfseries 06} (2017) 013}
  [\href{https://arxiv.org/abs/1701.08844}{{\ttfamily 1701.08844}}].

\bibitem{Balantekin:2007es}
A.~B. Balantekin, J.~Gava and C.~Volpe, \emph{{Possible CP-violation effects in
  core-collapse supernovae}},
  \href{https://doi.org/10.1016/j.physletb.2008.03.038}{\emph{Phys. Lett. B}
  {\bfseries 662} (2008) 396}
  [\href{https://arxiv.org/abs/0710.3112}{{\ttfamily 0710.3112}}].

\bibitem{Gava:2008rp}
J.~Gava and C.~Volpe, \emph{{Collective neutrino oscillations in matter and CP
  violation}}, \href{https://doi.org/10.1103/PhysRevD.78.083007}{\emph{Phys.
  Rev. D} {\bfseries 78} (2008) 083007}
  [\href{https://arxiv.org/abs/0807.3418}{{\ttfamily 0807.3418}}].

\end{thebibliography}\endgroup

\end{document}